\documentclass[preprint2]{aastex}

\def  \be   {\begin{equation}}
\def  \ee   {\end{equation}}
\def  \beq  {\begin{eqnarray}}
\def  \eeq  {\end{eqnarray}}
\def  \etal  {et~al.}

\shorttitle{Magnetic reconnection in partially ionized plasmas}
\shortauthors{Malyshkin \& Zweibel}

\begin{document}

\title{Onset of Fast Magnetic Reconnection in Partially Ionized Gases}

\author{Leonid M. Malyshkin\altaffilmark{1} and Ellen G. Zweibel\altaffilmark{2}}
\affil{${~}^1$Department of Astronomy \& Astrophysics,
University of Chicago, 5640 S. Ellis Ave., Chicago, IL 60637; {\sf leonmal@uchicago.edu}}
\affil{${~}^2$Departments of Astronomy and Physics, University of Wisconsin-Madison,  
6281 Chamberlain Hall, 475 N. Charter St., Madison, WI 53706; {\sf zweibel@astro.wisc.edu}}

%------------------------------------------------------------------------------------------

\begin{abstract}
We consider quasi-stationary two-dimensional magnetic reconnection 
in a partially ionized incompressible plasma. 
We find that when the plasma is weakly ionized and the collisions 
between the ions and the neutral particles are significant, 
the transition to fast collisionless reconnection due to the Hall 
effect in the generalized Ohm's law is expected to occur at 
much lower values of the Lundquist number, as compared to a fully 
ionized plasma case. 
We estimate that these conditions for fast reconnection are satisfied 
in molecular clouds and in protostellar disks.
\end{abstract}
\keywords{magnetic fields --- reconnection -- 
molecular clouds -- protostellar disks }

%------------------------------------------------------------------------------------------

\section{Introduction}

Magnetic reconnection plays a very 
important role in astrophysical plasmas. During the reconnection process magnetic 
energy is converted into plasma kinetic energy, thermal heat and acceleration 
of charged particles, and the topology of magnetic field lines is rearranged
\citep{kulsrud_2005,yamada_2010}. 
Magnetic reconnection is believed to be the power source behind various 
astrophysical phenomena, such as solar flares and geomagnetic storms. Magnetic
reconnection also frequently controls transport of charged particles and 
heat in interstellar and intergalactic media 
\citep{kulsrud_2005,zweibel_2009}.

In order for reconnection to be the energy release mechanism in transient phenomena 
such as solar flares, it must be fast
\citep{kulsrud_2005,uzdensky_2007,yamada_2010}.
Although slow reconnection is well explained 
by the Sweet-Parker model for reconnection in highly conductive, 
hot plasmas \citep{sweet_1958,parker_1963}, 
a common theoretical picture of fast magnetic 
reconnection has not emerged yet. A possible reason is that physical processes 
able to enhance dissipation in a reconnection layer and to cause fast 
reconnection are fairly complicated for a theoretical or experimental study. 
However, with development of supercomputers, considerable progress in 
understanding possible mechanisms of fast magnetic reconnection has been 
achieved by means of numerical simulations \citep{yamada_2010}.
In particular, one of the most important results that has been found both in 
simulations and in laboratory experiments is 
that in fully-ionized plasmas the transition from slow to fast reconnection 
occurs when the Sweet-Parker reconnection layer thickness becomes comparable 
to the ion inertial length, so that the Hall term in the generalized Ohm's law 
becomes important
\citep[for example, see][]{ma_bhattacharjee_1996, biskamp_1997, 
birn_etal_2001, cassak_2005, drake_2006, yamada_etal_2006}. Since this
condition is equivalent
to the collisional mean free path exceeding the length of the current sheet 
multiplied by $(m_e/m_i)^{1/2}$
\citep{zweibel_2009},
this type of reconnection is also referred to as fast collisionless reconnection.
Collisionless reconnection can occur in space plasmas, in the solar corona, and 
in hot accretion disks. 
It cannot occur in the interstellar medium, however, unless the reconnection layer 
is very short compared to macroscopic scales
(\citet{zweibel_2009}; see Section~\ref{SEC_DISCUSSION}).

In many astrophysical systems, such as much of the interstellar medium and the 
solar chromosphere, the ionization fraction is low. Studies of
{\textit{collisional}} reconnection in partially ionized gases have revealed 
two effects which bear on the reconnection process. If ion-neutral collisions 
are sufficiently weak, the plasma and
the neutrals decouple, so the reconnection speed is scaled by the Alfven speed
$V_{Ai}\equiv B/\sqrt{4\pi\rho_i}$ in the plasma alone, while in the strongly
collisional case the relevant speed is the bulk Alfven speed 
$V_A=B/\sqrt{4\pi\rho}$ \citep{zweibel_1989}. Thus, reconnection with weak 
friction is faster than reconnection with strong friction by $\sqrt{\rho/\rho_i}$. 
A separate effect is the thinning of magnetic 
neutral sheets \citep{brandenburg_1994}, which can dramatically
increase their merging rate \citep{zweibel_2003,lazarian_2004,hillier_2010}. 
It goes (almost) without saying
that neutrals also affect the reconnection process by making the plasma more 
resistive. This is an important effect in the low
chromosphere, in protostellar disks, and in the densest interstellar gas. 

The onset of collisionless, or Hall, reconnection in partially ionized gases has 
not yet been examined. 
In this paper we derive the condition for a transition to 
fast reconnection in partially ionized plasmas and apply our results to 
reconnection in molecular clouds, protostellar disks, and the solar chromosphere. 
We anticipate that flares in protostellar
disks will be observable with ALMA, and chromospheric flares with IRIS, making 
our results especially timely.

Our main results are as follows. When the Hall effect can be neglected, there 
are three regimes of reconnection in a
weakly ionized gas, which we 
refer to as weak, strong, and intermediate coupling, respectively. When the 
ion-neutral collision frequency $\nu_{in}$, reconnection current layer length 
$L$, and plasma Alfven speed $V_{Ai}$ satisfy
the inequality $\nu_{in}L/V_{Ai} < 1$, the neutrals are decoupled from the 
reconnection process, and the reconnection rate is
determined by the plasma parameters. When the neutral-ion collision frequency 
$\nu_{ni}=\nu_{in}\rho_i/\rho$, layer length $L$, 
and bulk Alfven
speed $V_A$ satisfy $\nu_{ni}L/V_A > 1$, the neutrals move with the plasma and 
the reconnection rate is
determined by the bulk parameters. In the third regime, the collisionality is 
intermediate, and dissipation by ion-neutral friction
is especially strong. The first two cases were described in \citet{zweibel_1989}, 
and all three are analogous to the regimes
of MHD wave propagation in partially ionized gases \citep{kulsrud_1969}. 

The onset of Hall reconnection occurs in the weakly
coupled regime just as it would for a fully ionized plasma (with resistivity 
modified by electron-neutral collisions). In the
strongly coupled regime, however, while the Sweet-Parker layer is thickened by 
enhanced resistivity and reduced effective Alfven
speed, the ion inertial scale can be
 increased even more, enlarging the parameter space for fast, Hall mediated 
reconnection. A similar
enhancement of the Hall effect in weakly ionized systems has been seen in 
studies of the magneto-rotational instability in
protostellar disks \citep{balbus_2001,salmeron_2005}.

In the next section we present basic three-fluid magnetohydrodynamics (MHD) 
equations for partially ionized plasmas. 
In Section~\ref{SEC_ASTRO_PARAMETERS} we discuss physical conditions in 
the interstellar medium (ISM).
In Section~\ref{SEC_RECONNECTION_EQUATIONS} we derive equations that 
describe quasi-stationary magnetic reconnection in partially ionized plasmas.
In Section~\ref{SEC_SOLUTION} we find the solution of the equations
and analyze it. 
Finally, in Section~\ref{SEC_DISCUSSION} we apply our results to magnetic 
reconnection in molecular clouds, protostellar disks, and the solar chromosphere.

As in many other studies of reconnection, we concentrate on the 2D case. We believe the three regimes of MHD reconnection
discussed here - corresponding to weak, strong, and intermediate ion-neutral coupling - are robust, as these have been found to
describe many other MHD phenomena in weakly ionized media. The increase in the ion - electron decoupling scale due to increased
effective ion mass is probably similarly robust. However, there are undoubtedly effects intrinsic to 3D which all 2D studies miss,
and our work shares these limitations.

%------------------------------------------------------------------------------------------

\section{Basic three-fluid MHD equations}
\label{SEC_BASIC_EQUATIONS}

In this paper, except in the next section, we use the Heaviside-Lorentz 
rationalized physical units, in which the speed of light $c$ and four 
times $\pi$ are replaced by unity, i.e.~$c\to 1$ and $4\pi\to 1$. 
In order to convert our equations to the Gaussian centimeter-gram-second 
(CGS) units, the following substitutions should be made: 
magnetic field ${\bf B}\rightarrow {\bf B}/\sqrt{4\pi}$, electric field 
${\bf E}\rightarrow c{\bf E}/\sqrt{4\pi}$, electric current 
${\bf j}\rightarrow\sqrt{4\pi}\,{\bf j}/c$, electrical resistivity 
$\eta\rightarrow\eta c^2/4\pi$, and the proton electric charge 
$e\rightarrow\sqrt{4\pi}\,e/c$.

Let us consider a partially ionized, non-relativistic, quasi-neutral, 
incompressible three-component plasma, which is composed of electrons, 
single-charged ions, and neutral particles. The momentum equations for 
these three components are~\citep{braginskii_1965,sturrock_1994} 
\begin{eqnarray}
&& 0 = -{\bf\nabla}P_e -ne({\bf E}+{\bf u}^e\times{\bf B})
\nonumber
\\
&& \quad {}-\rho_e\nu_{ei}({\bf u}^e-{\bf u}^i)
-\rho_e\nu_{en}({\bf u}^e-{\bf u}^n),
\label{MOTION_E_ORIGINAL}
\\
&& \rho_i\left[\partial_t{\bf u}^i+({\bf u}^i{\bf\nabla}){\bf u}^i\right] = 
-{\bf\nabla}P_i 
\nonumber
\\
&& \quad {}+ne({\bf E}+{\bf u}^i\times{\bf B})
\nonumber
\\
&& \quad {}+\rho_e\nu_{ei}({\bf u}^e-{\bf u}^i)
-\rho_i\nu_{in}({\bf u}^i-{\bf u}^n), \quad
\label{MOTION_I_ORIGINAL}
\\
&& \rho_n\left[\partial_t{\bf u}^n+({\bf u}^n{\bf\nabla}){\bf u}^n\right] = 
-{\bf\nabla}P_n 
\nonumber
\\
&& \quad {}+\rho_i\nu_{in}({\bf u}^i-{\bf u}^n)
+\rho_e\nu_{en}({\bf u}^e-{\bf u}^n).
\label{MOTION_N_ORIGINAL}
\end{eqnarray}
Here, for simplicity, we neglect electron inertia on the left-hand-side 
of eq.~(\ref{MOTION_E_ORIGINAL}); $n$ is the electron number density, 
equal to that of the ions in a quasi-neutral plasma;
${\bf u}$, $\rho$ and $P$ are velocity, mass density and pressure 
respectively; we assume the pressure tensors are well approximated by scalars for
all species. The subscripts and superscripts ``e'', ``i'' and ``n'' 
refer to the electrons, ions and neutral particles. 
The last two terms on the right-hand-side of each of the 
eqs.~(\ref{MOTION_E_ORIGINAL})-(\ref{MOTION_N_ORIGINAL}) represent 
the momentum exchange between the plasma components 
due to electron-ion, electron-neutral 
and ion-neutral collisions with effective frequencies $\nu_{ei}$, 
$\nu_{en}$ and $\nu_{in}$ respectively. For simplicity, we neglect 
electron-electron, ion-ion and neutral-neutral collisions and the 
corresponding viscous forces. Also, in this study we neglect ionization
and recombination processes, and, therefore, the densities 
\beq
\rho_e=nm_e, \quad
\rho_i=nm_i, \quad
\rho_n=n_nm_n
\label{RHO}
\eeq
of the three plasma components are constant in the 
incompressible plasma case. Here $m_e$, $m_i$ and $m_n$ are the 
electron, ion and neutral masses respectively, and $n_n$ is
the neutral number density.

The electric current is ${\bf j}=ne({\bf u}^i-{\bf u}^e)$,
and, therefore, the electron velocity is
\begin{eqnarray}\label{CURRENT}
{\bf u}^e={\bf u}^i-{\bf j}/ne.
\label{U_E}
\end{eqnarray}
Substituting eq. (\ref{CURRENT}) into eq.~(\ref{MOTION_E_ORIGINAL}), 
we obtain Ohm's law
\begin{eqnarray}
{\bf E} &\!\!=\!\!&
\eta\,{\bf j}-{\bf u}^i\times{\bf B}+(1/ne)\,{\bf j}\times{\bf B}
\nonumber
\\
&& 
{}-(1/ne){\bf\nabla}P_e -(m_e\nu_{en}/e)({\bf u}^i-{\bf u}^n),
\quad
\label{OHMS_LAW_ORIGINAL}
\end{eqnarray}
where ${\bf j}\times{\bf B}/ne$ is the Hall term, and 
\begin{eqnarray}
\begin{array}{l}
\eta=(\nu_{ei}+\nu_{en})m_e/ne^2=\eta_{ei}+\eta_{en},
\\
\eta_{ei}=\nu_{ei}m_e/ne^2=\nu_{ei}d_e^2, 
\\
\eta_{en}=\nu_{en}m_e/ne^2=\nu_{en}d_e^2,
\\
d_e\equiv(m_e/ne^2)^{1/2}.
\end{array}
\label{RESISTIVITY}
\end{eqnarray}
Here $\eta_{ei}$ is the standard Spitzer resistivity~\citep{sturrock_1994}, 
$\eta_{en}$ is the resistivity due to the electron-neutral collisions, 
$\eta$ is the total resistivity, and $d_e$ is the electron inertial length. 
The total resistivity $\eta$ is enhanced over the Spitzer value by the 
electron-neutral collisions, as one expects.

It will sometimes be useful to work with the electron collision 
time $\tau_e\equiv (\nu_{ei}+\nu_{en})^{-1}$. In terms of $\tau_e$,
\beq
\eta\equiv\frac{d_e^2}{\tau_e}.
\label{ETA_TAU_E}
\eeq

Next, we take the sum of 
equations~(\ref{MOTION_E_ORIGINAL})-(\ref{MOTION_I_ORIGINAL})
and use formula~(\ref{U_E}). As a result, we obtain the momentum
equation for the ions:
\begin{eqnarray}
&& {}\!\!\!\!\!\!\!\! 
\rho_i\left[\partial_t{\bf u}^i+({\bf u}^i{\bf\nabla}){\bf u}^i\right] = 
-{\bf\nabla}(P_e+P_i) + {\bf j}\times{\bf B}
\nonumber
\\
&& \!\!\!\!\! {}-(\rho_i\nu_{in}+\rho_e\nu_{en})({\bf u}^i-{\bf u}^n)
+(m_e\nu_{en}/e){\bf j},
\qquad
\label{MOMENTUM_I_ORIGINAL}
\end{eqnarray}
Finally, we substitute eq.~(\ref{U_E}) into eq.~(\ref{MOTION_N_ORIGINAL}) 
and rewrite the momentum equation for the neutrals as
\begin{eqnarray}
&& {}\!\!\!\!\!\!\!\! 
\rho_n\left[\partial_t{\bf u}^n+({\bf u}^n{\bf\nabla}){\bf u}^n\right] = 
-{\bf\nabla}P_n 
\nonumber
\\
&& \!\!\!\!\! {}+(\rho_i\nu_{in}+\rho_e\nu_{en})({\bf u}^i-{\bf u}^n)
-(m_e\nu_{en}/e){\bf j}.
\qquad
\label{MOMENTUM_N_ORIGINAL}
\end{eqnarray}

Equations~(\ref{OHMS_LAW_ORIGINAL})-(\ref{MOMENTUM_N_ORIGINAL}) 
together with the Maxwell equations are the basic three-fluid 
MHD equations for a partially ionized plasma. In addition to 
these equations, we 
note that in incompressible and non-relativistic plasmas the velocities 
and the electric current are divergence-free, 
${\bf\nabla}\cdot{\bf u}^i={\bf\nabla}\cdot{\bf u}^n=0$ 
and ${\bf\nabla}\cdot{\bf j}=0$.

%------------------------------------------------------------------------------------------

\section{Numerical Expressions for Parameters}
\label{SEC_ASTRO_PARAMETERS}

Let us estimate the values of physical parameters in representative 
astrophysical plasmas. This is useful to
motivate some approximations. In Section~\ref{SEC_DISCUSSION}, we will 
apply the theoretical results of this study to the weakly ionized interstellar 
medium (ISM), protostellar disks, and the solar chromosphere. In this section 
we temporarily use the Gaussian centimeter-gram-second (CGS) physical units.

The ion-neutral, electron-neutral and electron-ion collisional frequencies 
are \citep{braginskii_1965,draine_1983}
\beq
\!\!
\begin{array}{l}
\displaystyle
\nu_{in} \approx 1.9\!\times\! 10^{-9}sec^{-1}\,
\frac{\rho_n/\rho_i}{1+m_n/m_i}\;n,
\\
\displaystyle
\nu_{en} \approx 8.3\!\times\! 10^{-10}sec^{-1}\,
\frac{\rho_n}{\rho_i}\,\frac{m_i}{m_n}\;
n\,T_K^{1/2},
\\
\displaystyle
\nu_{ei}\, \approx 60\,sec^{-1}\:
n\,T_K^{-3/2}.
\end{array}
\!\!
\label{NU_VALUE}
\eeq
Here the frequencies are measured in inverse seconds (Hertz), 
the electron number density $n$ is measured in $cm^{-3}$, and 
the electron temperature $T$ is in Kelvins. 
We multiply the expression for $\nu_{in}$ by a factor of four in making 
estimates for the solar
chromosphere, due to its relatively large temperature \citep{de_pontieu_2001}.
The total electrical resistivity of the magnetic field, given by
eq.~(\ref{RESISTIVITY}), is relatively small,
\beq
\eta &\!\!\approx\!\!& 2.4\times 10^{-7}sec \times T_K^{-3/2}
\nonumber\\
&\!\!\!\!& 
\times \big[1+ 1.4\times 10^{-11}(\rho_n/\rho_i)(m_i/m_n)\,T_K^2\big].
\quad
\nonumber
\eeq
As a result, the characteristic Lundquist number 
$S_i=V_{Ai}L_{ext}\left/(\eta c^2/4\pi)\right.$ is very large 
in cosmic plasmas,
\beq
S_i &\!\!\!\approx\!\!\!& 2\times 10^5\,(m_p/m_i)^{1/2}
\nonumber\\
&\!\!\!\!\!\!& \times\frac{L_{ext,AU}\,B_{ext,\mu G}\;
T_K^{3/2}\,n_{cm^{-3}}^{-1/2}}
{1+ 1.4\!\times\! 10^{-11}(\rho_n/\rho_i)(m_i/m_n)\,T_K^2}\gg
\nonumber\\
&\!\!\!\gg\!\!\!&1.
\label{S_i_VALUE}
\eeq
Here $L_{ext,AU}$ is a characteristic system size in the astronomical 
units (AU), $B_{ext,\mu G}$ is the reconnecting magnetic field in 
microgauss ($\mu G$), $m_p$ is the proton mass, and
velocity $V_{Ai}=B_{ext}/\sqrt{4\pi nm_i}$ is the Alfven velocity 
based on the ion density,
\beq
V_{Ai}\approx 2.2\times 10^5\,
\frac{cm}{sec}\;\frac{m_p^{1/2}}{m_i^{1/2}}\;
\frac{B_{ext,\mu G}}{n_{cm^{-3}}^{1/2}}.
\label{V_Ai_VALUE}
\eeq

As we shall see below, it is useful to introduce the ion inertial 
length $d_i=(m_ic^2/4\pi ne^2)^{1/2}$ [in Heaviside-Lorentz units, 
$d_i=(m_i/ne^2)^{1/2}$; see the last of eqs.~(\ref{RESISTIVITY})]. 
Its approximate value is
\beq
d_i \approx 2.3\times 10^7 cm\;(m_i/m_p)^{1/2}\,n_{cm^{-3}}^{-1/2}.
\label{D_i_VALUE}
\eeq

Useful alternative expressions for $S_i$ and $V_{Ai}$ are 
\begin{equation}\label{Si}
S_i=\frac{L_{ext}}{d_i}\omega_{ce}\tau_e,
\qquad
V_{Ai}=\frac{m_e}{m_i}\omega_{ce}d_i,
\label{V_Ai_VALUE_2}
\end{equation}
where $\omega_{ce}=eB/m_ec$ is the electron cyclotron frequency.

Using eqs.~(\ref{RESISTIVITY}) and~(\ref{NU_VALUE}), let us estimate 
the following important dimensionless ratios, which show the relative 
strength of particle collisions:
\beq
\frac{m_e\nu_{en}}{m_i\nu_{in}} &\!\!\approx\!\!& 
0.00024\;T_K^{1/2}\!\left[\frac{m_p}{m_i}+\frac{m_p}{m_n}\right]\ll 1,
\quad
\label{NU_E_OVER_NU_I}
\\
\frac{\eta_{en}}{\eta_{ei}} &\!\!=\!\!& 
\frac{\nu_{en}}{\nu_{ei}}\approx 
1.4\!\times\! 10^{-11}\,\frac{\rho_n}{\rho_i}\,\frac{m_i}{m_n}\,T_K^2. 
\label{ETA_en_OVER_ETA}
\eeq
Equation~(\ref{NU_E_OVER_NU_I}) is an estimate for the ratio of the last 
two terms in eq.~(\ref{MOTION_N_ORIGINAL}). We see that it is very small, 
unless the electron temperature is several millions degrees. Thus, 
due to relatively small electron mass, the effect that the neutral particles 
experience from their collisions with the electrons is typically negligible 
as compared to the effect from the ion-neutral collisions. In contrast, 
the relative strength of the electron-ion and electron-neutral collisions,
given by eq.~(\ref{ETA_en_OVER_ETA}), can be either large or small, 
depending on the density ratio $\rho_n/\rho_i$ and on the electron 
temperature $T$. Therefore, the total resistivity $\eta$, 
given by eq.~(\ref{RESISTIVITY}), can be dominated by either 
electron-ion collisions or by electron-neutral collisions 
in interstellar medium and in laboratory plasma experiments.

%------------------------------------------------------------------------------------------

\section{Reconnection equations}
\label{SEC_RECONNECTION_EQUATIONS}

In this section let us derive equations that describe the magnetic 
reconnection process in partially ionized plasmas. 

It turns out that, when inequality 
$m_e\nu_{en}\ll m_i\nu_{in}$ holds in a system undergoing magnetic
reconnection [refer to eq.~(\ref{NU_E_OVER_NU_I})], the electron-neutral 
collisions can be neglected in all equations, except in 
eq.~(\ref{RESISTIVITY}) for the total resistivity. 
The proof is given in Appendix~\ref{APPENDIX_A}. 
As a result, we can omit the terms proportional to the 
electron-neutral collision frequency $\nu_{en}$ in 
eqs.~(\ref{OHMS_LAW_ORIGINAL}), (\ref{MOMENTUM_I_ORIGINAL}) 
and~(\ref{MOMENTUM_N_ORIGINAL}), and can rewrite these 
equations as
\beq
&& {}\!\!\!\!\!\!\!\! 
{\bf E} =
\eta\,{\bf j}-{\bf u}^i\times{\bf B}+{\bf j}\times{\bf B}/ne
-{\bf\nabla}P_e/ne,
\label{OHMS_LAW}
\\
&& {}\!\!\!\!\!\!\!\! 
\rho_i\left[\partial_t{\bf u}^i+({\bf u}^i{\bf\nabla}){\bf u}^i\right] = 
-{\bf\nabla}(P_e+P_i) + {\bf j}\times{\bf B}
\nonumber\\
&& {}\!\!\!\!\!\!\!\! 
\hphantom{\rho_i\left[\partial_t{\bf u}^i+({\bf u}^i{\bf\nabla}){\bf u}^i\right] = {}}
-\rho_i\nu_{in}({\bf u}^i-{\bf u}^n),
\qquad
\label{MOMENTUM_I}
\\
&& {}\!\!\!\!\!\!\!\! 
\rho_n\left[\partial_t{\bf u}^n+({\bf u}^n{\bf\nabla}){\bf u}^n\right] = 
-{\bf\nabla}P_n 
\nonumber\\
&& {}\!\!\!\!\!\!\!\! 
\hphantom{\rho_n\left[\partial_t{\bf u}^n+({\bf u}^n{\bf\nabla}){\bf u}^n\right] = {}}
+\rho_i\nu_{in}({\bf u}^i-{\bf u}^n),
\qquad
\label{MOMENTUM_N}
\end{eqnarray}
where the total resistivity $\eta$ is given by eq.~(\ref{RESISTIVITY})
and includes a contribution from electron-neutral collisions.

\begin{figure}[t]
\vspace{3.5truecm}
\includegraphics{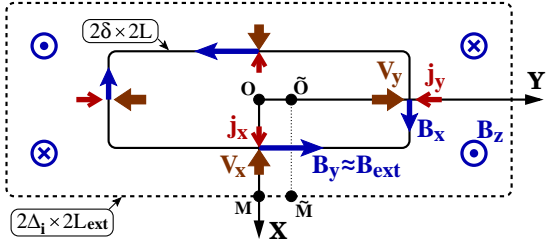}
\caption{The geometry of the reconnection layer, with common notations. 
The point $O$ is a magnetic $X$-point. The fluid flows toward $O$ along
the $x$ axis and away from $O$ along the $y$ axis, carrying the 
magnetic field. The field is frozen into the electron fluid
everywhere except inside the reconnection current layer, which has width $2\delta$ 
and length $2L$. In MHD reconnection the ion and electron decoupling regions
coincide, but in Hall reconnection the ion decoupling layer is larger; its 
width and length are $2\Delta_i$ and $2L_{ext}$, respectively. The points
$\tilde O$, $M$, and $\tilde M$ are defined in Appendix B. For additional 
explanation, see the text.
\label{FIGURE_LAYER}
}
\end{figure}

Let us now describe the reconnection layer, shown 
in Figure~\ref{FIGURE_LAYER}. We assume the classical two-dimensional 
Sweet-Parker-Petschek geometry for the reconnection layer. The 
layer lies in the $x$-$y$ plane of the coordinate system, 
and the $x$- and $y$-axes are chosen to be perpendicular to 
and along the reconnection layer respectively. The $z$ derivatives 
of all physical quantities are assumed to be zero. 

The thickness of the reconnection current layer is $2\delta$, which 
can be formally defined by fitting the Harris sheet profile 
$(B_{ext}/\delta)cosh^{-2}(x/\delta)$ to the current profile 
$j_z(x,y=0)$. The length of the reconnection current layer is $2L$. 
Outside the reconnection current layer the z-component of the  
Ohm's law~(\ref{OHMS_LAW}) reduces to $E_z=-({\bf u}^e\times{\bf B})_z$ 
[see also eq.~(\ref{U_E})], and, therefore, the magnetic field lines 
are frozen into the electron fluid. Thus, the reconnection 
current layer coincides with the electron layer, which is the region 
where the electrons are decoupled from the field lines. 

The ion layer, which is the region where the ions are decoupled from the 
field lines, can be much larger. We use notations $2\Delta_i$ and $2L_{ext}$ 
for the ion layer thickness and length, where $L_{ext}$ is also 
approximately equal to the external (global) scale of the magnetic field. 
We have $\Delta_i\gtrsim\delta$ and $L_{ext}\gtrsim L$. The region
where the neutral particles are decoupled from the ions can be still 
larger than the ion layer. 

The value of the reconnecting field $B_y$ in the upstream regions outside 
the reconnection layer (at $x\approx\delta$) is approximately equal to  
the value of the external (global) magnetic field $B_{ext}$ outside 
the ion layer, up to a factor of order unity. This can easily 
be seen from the definition of $\delta$ and from the Ampere's law 
$z$-component $B_y(x,y=0)\approx\int_0^x j_z(x',y=0)dx'$.
The out-of-plane field $B_z$ is assumed to have a quadrupole 
structure~\citep{drake_2006,hesse_2007,yamada_2010,zweibel_2009}.
Finally, the reconnection layer is assumed to have a point symmetry with 
respect to its geometric center, point~$O$ shown of Figure~\ref{FIGURE_LAYER}. 
As a result of reflection symmetries with respect to the $x$- and $y$-axes, 
the $x$-, $y$- and $z$-components of ${\bf u}$, ${\bf B}$ and 
${\bf j}$ have the following symmetries:
$u_x(\pm x,\mp y)=\pm u_x(x,y)$, $u_y(\pm x,\mp y)=\mp u_y(x,y)$,
$u_z(\pm x,\mp y)=u_z(x,y)$,
$B_x(\pm x,\mp y)=\mp B_x(x,y)$, $B_y(\pm x,\mp y)=\pm B_y(x,y)$,
$B_z(\pm x,\mp y)=-B_z(x,y)$,
$j_x(\pm x,\mp y)=\pm j_x(x,y)$, $j_y(\pm x,\mp y)=\mp j_y(x,y)$
and $j_z(\pm x,\mp y)=j_z(x,y)$. Here ${\bf u}$ is the velocity of any species.
We extensively use these symmetries 
in the forthcoming analytical derivations, which are similar to the 
derivations in \citet{malyshkin_2008}. 

Let us list the assumptions that we make for the reconnection process 
in a partially ionized plasma. First, as we have already stated above, we
neglect ionization and recombination processes. 
Second, we assume that the collision frequencies and resistivities 
$\eta$, $\eta_{ei}$, $\eta_{en}$ are constant in space and time. 
We also assume that the characteristic Lundquist number $S_i$ is 
very large,
\beq
S_i = V_{Ai}L_{ext}/\eta \gg 1, \quad V_{Ai} = B_{ext}/\sqrt{\rho_i},
\label{Si_VAi}
\eeq
an assumption easily satisfied in cosmic plasmas [see eq.~(\ref{S_i_VALUE})].
Note that the Alfven velocity $V_{Ai}$ is calculated by using the ion 
density $\rho_i$ and the reconnecting magnetic field value $B_{ext}$. 
Third, we assume that the reconnection process is stationary or 
quasi-stationary, so that all time derivatives can be neglected in all
equations. This assumption means that the reconnection 
rate is slow sub-Alfvenic, $E_z\ll V_{Ai}B_{ext}$, and that there are no 
plasma instabilities in the reconnection layer. 
Fourth, we assume that the reconnection layer is thin, $\delta\ll L$ and
$\Delta_i\ll L$. 
This assumption is related to the previous assumption of slow reconnection 
because of the mass conservation condition for the plasma.

Before we proceed with derivations of the reconnection rate it is convenient
to introduce the following dimensionless parameters:
\beq
{\tilde\rho} &\!\!\equiv\!\!& \rho_n/\rho_i,
\label{DENSITY_NORM}
\\
{\tilde\nu} &\!\!\equiv\!\!& 
\nu_{in}\left/2(\partial_y u_y^i)_o \right. ,
\label{NU_NORM}
\\
{\tilde\upsilon} &\!\!\equiv\!\!& 
(\partial_y u_y^n)_o\left/(\partial_y u_y^i)_o \right. ,
\label{U_N_NORM}
\\
{\tilde\gamma} &\!\!\equiv\!\!& 
(\partial_{xy}B_z)_o\left/ne(\partial_y u_y^i)_o \right. .
\label{GAMMA}
\eeq
In eq. (\ref{DENSITY_NORM}), ${\tilde\rho}$ is the ratio of the 
densities of the neutrals and 
the ions. Equation~(\ref{NU_NORM}) introduces the ion-neutral collision 
frequency, normalized by two times $(\partial_y u_y^i)_o$. 
The latter is the ion acceleration rate $\partial_y u_y^i$ calculated 
at the central point~$O$ (see Figure~\ref{FIGURE_LAYER}).  
The parameter ${\tilde\upsilon}$, defined by eq.~(\ref{U_N_NORM}), is 
approximately the ratio of the acceleration rates of the 
outflowing neutrals and the ions inside the reconnection current layer. 
Finally, the parameter ${\tilde\gamma}$, defined by eq.~(\ref{GAMMA}), is 
the normalized value of $(\partial_{xy}B_z)_o$ that is the second order 
mixed derivative of the quadrupole out-of-plane field $B_z$ at the central 
point~$O$. This parameter ${\tilde\gamma}$ gives the approximate ratio of 
the Hall term ${\bf j}\times{\bf B}/ne$ and the $-{\bf u}^i\times{\bf B}$ 
term that enter Ohm's law~(\ref{OHMS_LAW}), inside the reconnection 
current layer. Thus, the Hall term is important when 
${\tilde\gamma}\gtrsim 1$. Note that all 
parameters~(\ref{DENSITY_NORM})-(\ref{GAMMA}) are non-negative. We will 
often replace derivatives by inverse
length scales in estimating these parameters.

Now let us use eqs.~(\ref{OHMS_LAW})-(\ref{MOMENTUM_N}), the Maxwell 
equations and the incompressibility relations
\beq
\partial_x u_x^i=-\partial_y u_y^i, 
\qquad
\partial_x u_x^n=-\partial_y u_y^n, 
\label{DIV_U}
\eeq
to derive the formulas that we will later solve for the reconnection 
rate and other physical quantities.

First, we use Ampere's law. The displacement current can be neglected 
in a non-relativistic plasma, therefore, we find
\beq
\begin{array}{l}
j_x=\partial_y B_z, \qquad
j_y=-\partial_x B_z, 
\\
j_z=\partial_x B_y-\partial_y B_x.
\end{array}
\label{J_COMPONENTS}
\eeq
We can estimate the $z$-component of the electric current, $j_z$, at 
the central point~$O$ as
\begin{eqnarray}
j_o &\!\!\equiv\!\!& (j_z)_o=(\partial_x B_y-\partial_y B_x)_o \approx{}
\nonumber
\\
{} &\!\!\approx\!\!& (\partial_x B_y)_o \approx B_{ext}/\delta,
\label{AMPERES_LAW}
\end{eqnarray}
where we use $(\partial_y B_x)_o\ll (\partial_x B_y)_o\approx B_{ext}/\delta$
at the point~$O$. The last estimate, $(\partial_x B_y)_o\approx B_{ext}/\delta$, 
follows from the fact that $\delta$ is defined as the half-thickness of
the $j_z$ profile across the reconnection layer.

Faraday's law ${{\bf\nabla}\times{\bf E}}=-\partial_t{\bf B}$ for
the $x$- and $y$-components of the magnetic field gives
$\partial_y E_z=-\partial_t B_x=0$ and $\partial_x E_z=\partial_t B_y=0$,
where the time derivatives are neglected in the case of  quasi-stationary 
reconnection. As a result, the z-component of the electric field $E_z$ is 
constant in space, 
\begin{eqnarray}
E_z &\!\!=\!\!& \eta j_z-(u_x^i-j_x/ne)B_y+(u_y^i-j_y/ne)B_x
\nonumber\\
&\!\!=\!\!& \mbox{constant}.
\label{OHMS_LAW_Z}
\end{eqnarray}
Here we use Ohm's law~(\ref{OHMS_LAW}) to find the expression for $E_z$. 
The reconnection rate is the rate of destruction of the magnetic flux,
${}-\partial_t\int_0^\infty B_y\,dx=-\int_0^\infty \partial_t B_y\,dx
=-\int_0^\infty \partial_x E_z\,dx=(E_z)_o$. Thus, it is given by the 
value of $E_z$ at the central point~$O$,
\beq
E_z=\eta j_o.
\label{OHMS_LAW_Z_CENTER}
\eeq
One of our goals is to find the value of the reconnection current $j_o$ 
and to calculate the reconnection rate given by 
eq.~(\ref{OHMS_LAW_Z_CENTER}). 

With the time derivatives neglected, the $z$-components of the momentum 
equations~(\ref{MOMENTUM_I}) and~(\ref{MOMENTUM_N}) are
\beq
\rho_i(u_x^i\partial_x u_z^i+u_y^i\partial_y u_z^i) &\!\!\!=\!\!\!& 
j_xB_y-j_yB_x
\nonumber
\\
&\!\!\!\!\!\!& -\rho_i\nu_{in}(u_z^i-u_z^n),
\qquad
\label{MOMENTUM_I_Z}
\\
\rho_n(u_x^n\partial_x u_z^n+u_y^n\partial_y u_z^n) &\!\!\!=\!\!\!& 
\rho_i\nu_{in}(u_z^i-u_z^n).
\label{MOMENTUM_N_Z}
\end{eqnarray}
Calculating $\partial^2/\partial x^2$ 
of these equations at the central point~$O$, dividing the resulting 
expressions by $2\rho_i(\partial_y u_y^i)_o$, and using 
eqs.~(\ref{DENSITY_NORM})-(\ref{J_COMPONENTS}), we obtain
\beq
&& (1-{\tilde\nu})(\partial_{xx} u_z^i)_o 
+ {\tilde\nu}(\partial_{xx} u_z^n)_o = {}
\nonumber\\
&& \qquad\qquad\qquad\qquad 
{} = -{\tilde\gamma}(e/m_i)(\partial_x B_y)_o, 
\qquad
\label{Uz_I_XX_EQ}
\\
&& {\tilde\nu}(\partial_{xx} u_z^i)_o 
+ ({\tilde\rho}{\tilde\upsilon}-{\tilde\nu})(\partial_{xx} u_z^n)_o = 0.
\label{Uz_N_XX_EQ}
\eeq
Similarly, taking $\partial^2/\partial y^2$ 
of equations~(\ref{MOMENTUM_I_Z}) and~(\ref{MOMENTUM_N_Z}) at 
the point~$O$, we find
\beq
&& (1+{\tilde\nu})(\partial_{yy} u_z^i)_o 
- {\tilde\nu}(\partial_{yy} u_z^n)_o = {}
\nonumber\\
&& \qquad\qquad\qquad\qquad 
{} = {\tilde\gamma}(e/m_i)(\partial_y B_x)_o,
\qquad
\label{Uz_I_YY_EQ}
\\
&& {}-{\tilde\nu}(\partial_{yy} u_z^i)_o 
+ ({\tilde\rho}{\tilde\upsilon}+{\tilde\nu})(\partial_{yy} u_z^n)_o = 0.
\label{Uz_N_YY_EQ}
\eeq
Equations~(\ref{Uz_I_XX_EQ})-(\ref{Uz_N_YY_EQ}) are two systems of 
two linear equations in each system for the unknown quantities 
$(\partial_{xx} u_z^i)_o$, $(\partial_{xx} u_z^n)_o$, 
$(\partial_{yy} u_z^i)_o$ and $(\partial_{yy} u_z^n)_o$. 
The solution is
\beq
(\partial_{xx} u_z^i)_o &\!\!\!\!=\!\!\!\!&
-{\tilde\gamma}(1+{\tilde\nu}{\tilde\rho}{\tilde\upsilon}/D_-)
(e/m_i)(\partial_x B_y)_o, 
\quad\;
\label{Uz_I_XX}
\\
(\partial_{xx} u_z^n)_o &\!\!\!\!=\!\!\!\!&
{\tilde\gamma}({\tilde\nu}/D_-)(e/m_i)(\partial_x B_y)_o,
\label{Uz_N_XX}
\\
(\partial_{yy} u_z^i)_o &\!\!\!\!=\!\!\!\!&
{\tilde\gamma}(1-{\tilde\nu}{\tilde\rho}{\tilde\upsilon}/D_+)
(e/m_i)(\partial_y B_x)_o, 
\label{Uz_I_YY}
\\
(\partial_{yy} u_z^n)_o &\!\!\!\!=\!\!\!\!&
{\tilde\gamma}({\tilde\nu}/D_+)(e/m_i)(\partial_y B_x)_o,
\label{Uz_N_YY}
\\
D_\pm &\!\!\!\!\equiv\!\!\!\!& {\tilde\rho}{\tilde\upsilon}
\pm{\tilde\nu}(1+{\tilde\rho}{\tilde\upsilon}).
\label{D}
\eeq
The determinant $D_+$ of system~(\ref{Uz_I_YY_EQ})-(\ref{Uz_N_YY_EQ}) 
is always positive, except for the trivial case when there are no 
collisions with the neutral particles (${\tilde\nu}=0$). 
At the same time, the determinant $D_-$ 
of system~(\ref{Uz_I_XX_EQ})-(\ref{Uz_N_XX_EQ}) can be zero 
even when ${\tilde\nu}\ne 0$. For now we will assume that $D_-$ is 
non-zero, the opposite case will be discussed 
below.\footnote{\label{D_MINUS_FOOTNOTE}
We shall see that the assumption $D_-\ne 0$ is satisfied if 
${\tilde\nu}\ll 1$ or if ${\tilde\nu}\gg 1$, the case 
${\tilde\nu}\approx 1$ will be considered separately.
} 

Next, we calculate the second-order derivatives $\partial^2/\partial x^2$
and $\partial^2/\partial y^2$ of equation~(\ref{OHMS_LAW_Z}) at 
the central point~$O$, using the fact that $E_z$ is constant. 
We find
\beq
0 &\!\!\!=\!\!\!& \eta(\partial_{xx}j_z)_o
-2[(\partial_x u_x^i)_o-(\partial_x j_x)_o/ne](\partial_x B_y)_o.
\nonumber
\\
0 &\!\!\!=\!\!\!& \eta(\partial_{yy}j_z)_o
+2[(\partial_y u_y^i)_o-(\partial_y j_y)_o/ne](\partial_y B_x)_o.
\nonumber
\eeq
We rewrite these formulas by using eqs.~(\ref{RESISTIVITY}), 
(\ref{DENSITY_NORM})-(\ref{J_COMPONENTS}) 
and~(\ref{Uz_I_XX})-(\ref{Uz_N_YY}). We obtain
\beq
{} -\eta(\partial_{xx}j_z)_o &\!\!=\!\!& 
2(\partial_y u_y^i)_o(\partial_x B_y)_o(1+{\tilde\gamma}),
\label{Ez_XX_INITIAL}
\\
{} -\eta(\partial_{yy}j_z)_o &\!\!=\!\!& 
2(\partial_y u_y^i)_o(\partial_y B_x)_o(1+{\tilde\gamma}).
\qquad
\label{Ez_YY_INITIAL}
\eeq
Taking the ratio of these two equations, we find
\beq
(\partial_y B_x)_o = (\partial_x B_y)_o
\frac{(\partial_{yy}j_z)_o}{(\partial_{xx}j_z)_o}
\approx \frac{B_{ext}\delta}{L^2}
\approx \frac{B_{ext}^2}{L^2j_o},
\label{Bx_Y}
\eeq
where we use eq.~(\ref{AMPERES_LAW}) and the estimates 
$(\partial_{yy}j_z)_o\approx -j_o/L^2$ and 
$(\partial_{xx}j_z)_o\approx -j_o/\delta^2$. We again use 
these estimates and eq.~(\ref{AMPERES_LAW}), to rewrite 
equation~(\ref{Ez_XX_INITIAL}) as
\beq
\eta j_o^2 \approx
B_{ext}^2(\partial_y u_y^i)_o(1+{\tilde\gamma}).
\qquad
\label{Ez_XX}
\eeq
Here and below we neglect all factors of order unity.
It is noteworthy that equation~(\ref{Ez_XX}) describes the supply 
of magnetic energy $B_{ext}^2$ into the reconnection layer, where it 
is dissipated by the Joule heating $\eta j_o^2$. The 
rate of magnetic energy supply, 
$(\partial_y u_y^i)_o(1+{\tilde\gamma})=
(\partial_y u_y^i)_o-(\partial_y j_y)_o/ne=(\partial_y u_y^e)_o$,
is equal to the electron velocity derivative 
because magnetic field lines are frozen into the electron fluid 
outside the reconnection current layer.

Now we use Faraday's law ${{\bf\nabla}\times{\bf E}}=-\partial_t{\bf B}$ for
the $z$-component of the magnetic field. We have 
$\partial_x E_y-\partial_y E_x=-\partial_t B_z=0$, where the time derivative \
is neglected again. We substitute $E_x$ and $E_y$ into this formula 
from Ohm's law~(\ref{OHMS_LAW}) and obtain
\beq
0 &\!\!=\!\!& \eta(\partial_x j_y-\partial_y j_x)
+(B_x\partial_x j_z+B_y\partial_y j_z)/ne
\nonumber
\\
&\!\!\!\!&
{}+u_x^i\partial_x B_z+u_y^i\partial_y B_z-B_x\partial_x u_z^i-B_y\partial_y u_z^i.
\nonumber
\eeq
Calculating the $\partial^2/\partial x\partial y$ derivative of this equation at 
the central point~$O$ and using equations~(\ref{DIV_U}), (\ref{J_COMPONENTS}), 
(\ref{Uz_I_XX}) and~(\ref{Uz_I_YY}), we obtain 
\begin{eqnarray}
0 &\!\!\!=\!\!\!&
-\eta\left[(\partial_{xx}+\partial_{yy})(\partial_{xy}B_z)\right]_o
\nonumber\\
&\!\!\!\!& +[(\partial_{xx}j_z)_o(\partial_y B_x)_o
+(\partial_{yy}j_z)_o(\partial_x B_y)_o]/ne
\nonumber\\
&\!\!\!\!& + (2{\tilde\gamma}{\tilde\nu}{\tilde\rho}^2{\tilde\upsilon}^2/D_-D_+)
(e/m_i)(\partial_x B_y)_o(\partial_y B_x)_o 
\nonumber
\\
&\!\!\!\approx\!\!\!& 
\eta{\tilde\gamma}ne(\partial_y u_y^i)_o/\delta^2
\nonumber\\
&\!\!\!\!& -[(j_o/\delta^2)(\partial_y B_x)_o+(j_o^2/L^2)]/ne
\nonumber\\
&\!\!\!\!& + ({\tilde\gamma}{\tilde\nu}{\tilde\rho}^2{\tilde\upsilon}^2/D_-D_+)
(e/m_i)j_o(\partial_y B_x)_o.
\label{Ez_XY_INITIAL}
\end{eqnarray}
To derive the final approximate expression, we use the estimates
$(\partial_{yy})_o\approx-1/L^2\ll(\partial_{xx})_o\approx-1/\delta^2$ and
$(\partial_x B_y)_o\approx j_o$, we also use 
equations~(\ref{RESISTIVITY}), (\ref{DENSITY_NORM})-(\ref{GAMMA}) and 
we drop factors of order unity. In Appendix~\ref{APPENDIX_A} we show that 
the last term in eq.~(\ref{Ez_XY_INITIAL}) can be neglected
(assuming $D_-\ne 0$). As a result, dropping this term and using 
eqs.~(\ref{Si_VAi}), (\ref{AMPERES_LAW}), (\ref{Bx_Y}), we obtain
\beq
{\tilde\gamma} &\!\!\approx\!\!& 
\frac{B_{ext}^2}{\eta\,n^2e^2L^2(\partial_y u_y^i)_o}.
\label{Ez_XY}
\eeq

Next, we consider the acceleration of the plasma in the $y$-direction, 
along the reconnection layer. We calculate $\partial/\partial y$ of 
the y-components of the momentum equations~(\ref{MOMENTUM_I}) 
and~(\ref{MOMENTUM_N}) at the central point~$O$, and, neglecting the 
time derivatives for a quasi-stationary reconnection, we obtain
\beq
\rho_i(\partial_y u_y^i)_o^2 &\!\!=\!\!& 
-(\partial_{yy}[P_e+P_i])_o +j_o(\partial_y B_x)_o
\nonumber\\
&\!\!\!\!& 
{}-\rho_i\nu_{in}(\partial_y u_y^i\!-\!\partial_y u_y^n)_o,
\label{Up_Y_Y_INITIAL}
\\
\rho_n(\partial_y u_y^n)_o^2 &\!\!=\!\!& 
-(\partial_{yy} P_n)_o
\nonumber\\
&\!\!\!\!& 
{}+\rho_i\nu_{in}(\partial_y u_y^i\!-\!\partial_y u_y^n)_o,
\qquad
\label{Un_Y_Y_INITIAL}
\eeq
In Appendix~{\ref{APPENDIX_B} we estimate the pressure terms 
and find that
\beq
&& {}\!\!\!\!\!\!\!\!\!\!\!\! 
(\partial_{yy}[P_e+P_i])_o \approx
-B_{ext}^2/L^2 +o\big\{j_o(\partial_yB_x)_o\big\}
\nonumber\\
&& {}\!\!\!\!\!\!\!\!\!\!\!\! 
\quad 
+o\big\{\rho_i(\partial_y u_y^i)_o^2\big\}
+o\big\{\rho_i\nu_{in}(\partial_y u_y^i\!-\!\partial_y u_y^n)_o\big\}, 
\quad\;\; 
\label{Pe_Pp_YY}
\\
&& {}\!\!\!\!\!\!\!\!\!\!\!\! 
(\partial_{yy} P_n)_o =
o\big\{\rho_n(\partial_y u_y^n)_o^2\big\}
\nonumber\\
&& {}\!\!\!\!\!\!\!\!\!\!\!\! 
\qquad \qquad \quad
+o\big\{\rho_i\nu_{in}(\partial_y u_y^i\!-\!\partial_y u_y^n)_o\big\},  
\label{Pn_YY}
\eeq
where $o\{...\}$ denotes terms that are small compared to the
expression inside the brackets $\{...\}$ in the case of a thin 
reconnection layer ($\delta\ll L$ and $\Delta_i\ll L$).
We substitute eqs.~(\ref{Pe_Pp_YY}) and~(\ref{Pn_YY}) 
into eqs.~(\ref{Up_Y_Y_INITIAL}) and~(\ref{Un_Y_Y_INITIAL}) 
and then use formulas~(\ref{DENSITY_NORM})-(\ref{GAMMA}), 
(\ref{J_COMPONENTS}). As a result, equation~(\ref{Un_Y_Y_INITIAL}) 
becomes
\beq
{\tilde\rho}{\tilde\upsilon}^2 = 2{\tilde\nu}(1-{\tilde\upsilon}),
\label{UPSILON_EQ}
\eeq
while the sum of eqs.~(\ref{Up_Y_Y_INITIAL}) and~(\ref{Un_Y_Y_INITIAL})
gives
\beq
(\partial_y u_y^i)_o^2\,(1+{\tilde\rho}{\tilde\upsilon}^2) \approx
V_{Ai}^2/L^2,
\label{Up_Y_Y_EQ}
\eeq
where we use eqs.~(\ref{Si_VAi}) and~(\ref{Bx_Y}).
Equation~(\ref{Up_Y_Y_EQ}) describes the increase of the total kinetic 
energy of the ions and neutrals due to the work produced by the 
pressure and magnetic forces during the plasma acceleration in the 
downstream regions. Note that parameter ${\tilde\upsilon}$, given 
by eq.~(\ref{U_N_NORM}), must be non-negative (to be more precise,
$0\le {\tilde\upsilon}\le 1$ must hold) because the neutral 
particles are dragged by collisions with the ions  
[see eq.~(\ref{MOMENTUM_N})]. Therefore, the physically correct solution
for ${\tilde\upsilon}$ of quadratic equation~(\ref{UPSILON_EQ}) is
\beq
{\tilde\upsilon}=
\big(\sqrt{{\tilde\nu}^2
+2{\tilde\rho}{\tilde\nu}}-{\tilde\nu}\big)/{\tilde\rho}
\,\,\approx\, \min\big\{ \sqrt{{\tilde\nu}/{\tilde\rho}},\, 1 \big\},
\label{UPSILON}
\eeq
where the final expression is a convenient simple estimate for 
${\tilde\upsilon}$.

Let us now estimate the thickness $\Delta_i$ of the ion layer 
(see Figure~\ref{FIGURE_LAYER}). Note that in the upstream region 
just outside the ion layer, at $x\approx\Delta_i$ and $y=0$, the 
electrons and ions are coupled together, the electric current is 
weak, the magnetic field lines are frozen into the electron-ion 
fluid, and eq.~(\ref{OHMS_LAW_Z}) reduces to 
$E_z = -u_x^i B_y\approx -u_x^i B_{ext}$, where 
$u_x^i\approx (\partial_x u_x^i)_o\Delta_i
=-(\partial_y u_y^i)_o\Delta_i$. Thus, we have
\beq
\Delta_i &\!\!\approx\!\!& E_z\big/(\partial_y u_y^i)_oB_{ext},
\label{DELTA_i}
\\
V_R &\!\!\approx\!\!& (\partial_y u_y^i)_o\Delta_i 
\approx E_z\big/B_{ext},
\label{V_R}
\eeq
where $V_R=|u_x^i|\approx(\partial_y u_y^i)_o\Delta_i$
is the reconnection velocity. It is the velocity with which magnetic 
field lines and magnetic energy are carried by the plasma 
into the reconnection region. 

Next, let us consider the thickness $\Delta_n$ of the region where 
the neutral particles are decoupled from the ions. If the 
neutrals and ions strongly collide and move together, 
${\tilde\upsilon}=1$, they are coupled everywhere, and 
$\Delta_n$ is not defined. If the neutrals 
and ions are not fully coupled and ${\tilde\upsilon}<1$,
a reasonable definition of $\Delta_n$ is based on the 
location upstream where the inflow velocities of the neutrals 
and ions become comparable. Namely, $u_x^n\approx u_x^i$ at 
$x\approx\Delta_n$ and $y=0$. Unfortunately, we cannot estimate
$\Delta_n$ defined this way by using our local, analytical approach.
This is because the profile of the neutrals inflow velocity $u_x^n$ 
as a function of $x$ is unknown in the upstream region outside 
the ion layer (i.e.~at $x\in[\Delta_i,\Delta_n]$).\footnote{
Note that if the ions and neutrals are weakly coupled and 
if the ion pressure force can be neglected, then the ion 
velocity $u_x^i\propto x^{-1/3}$ at $x\in[\Delta_i,\Delta_n]$
\citep{zweibel_2003}. However, the neutrals pressure force cannot 
be neglected because otherwise the inflow velocity of the neutrals 
would be much larger than eq.~(\ref{UPSILON}) implies
(to see this, integrate the x-component of eq.~(\ref{MOMENTUM_N}) 
over $x\in[\Delta_i,\Delta_n]$ at $y=0$, and use 
$\Delta_n\gg\Delta_i$). 
}
Finding this profile requires full solution of the governing PDEs.
Instead, we suggest a simple estimate $\Delta_n$, based on a
dimensional analysis, as follows. First, let us note that
the effective collision frequency for the neutrals is 
$\nu_{ni}=\nu_{in}/{\tilde\rho}$, which is obtained by comparing 
the first and the last terms in eq.~(\ref{MOMENTUM_N}). 
Second, the neutrals achieve their maximal inflow 
velocity around the edge of the ion layer, 
$|u_x^n(\Delta_i,0)|\approx |(\partial_x u_x^n)_o|\Delta_i
=(\partial_y u_y^n)_o\Delta_i
={\tilde\upsilon}(\partial_y u_y^i)_o\Delta_i$ 
[see eqs.~(\ref{U_N_NORM}) and~(\ref{DIV_U})].
Now, we can make an estimate
$\Delta_n\approx |u_x^n(\Delta_i,0)|/\nu_{ni}
\approx ({\tilde\rho}{\tilde\upsilon}/{\tilde\nu})\Delta_i
\approx \Delta_i\sqrt{{\tilde\rho}/{\tilde\nu}}
\approx \Delta_i/{\tilde\upsilon}\gtrsim \Delta_i$ 
in case ${\tilde\upsilon}\lesssim 1$ [see eqs.~(\ref{NU_NORM})
and~(\ref{UPSILON})]. 
Fortunately, the exact value of $\Delta_n$ does not directly 
influence the reconnection rate and other important physical 
parameters, calculated below.

In the end of this section let us estimate the energy 
dissipation rate due to the ion-neutral collisions, which 
heat the ions and the neutrals. The dissipation rate per unit 
time, per unit volume is 
$q_{in}=\rho_i\nu_{in}({\bf u}^i-{\bf u}^n)^2$.~\footnote{
To derive this formula, add together eq.~(\ref{MOMENTUM_I}) 
multiplied by ${\bf u}^i$ and eq.~(\ref{MOMENTUM_N}) 
multiplied by ${\bf u}^n$.
}
Therefore, the total dissipation (per unit time, per unit length in 
the $z$-direction) inside the upper right quarter of the ion layer 
is $Q_{in}=\int_0^L\int_0^{\Delta_i}q_{in}\,dx\,dy$.
The flux of the (electro)magnetic energy supplied into the 
ion layer is given by the x-component of the Poynting vector, 
$({\bf E}\times{\bf B})_x$. Therefore, the total magnetic energy 
supplied per unit time, per unit length in the $z$-direction, is 
${\cal E}_m\approx L|{\bf E}\times{\bf B}|_x
\approx LE_zB_{ext}$.
The ratio of the dissipated and supplied energy rates is
\beq
\frac{Q_{in}}{{\cal E}_m} &\!\!\!\approx\!\!\!&
\frac{\rho_i\nu_{in}}{LE_zB_{ext}}
\int_0^L\!\!\!\int_0^{\Delta_i}\! ({\bf u}^i-{\bf u}^n)^2 dx\,dy
\nonumber\\
&\!\!\!\approx\!\!\!& 
\frac{\rho_i\nu_{in}}{LE_zB_{ext}}
\bigg[L\!\int_0^{\Delta_i}\!(u_x^i-u_x^n)^2 dx
\nonumber\\
&\!\!\!\!\!\!& 
\hphantom{ \frac{\rho_i\nu_{in}}{LE_zB_{ext}}\bigg[ }
{}+\Delta_i\!\int_0^L\!(u_y^i-u_y^n)^2 dy\bigg]
\nonumber\\
&\!\!\!\approx\!\!\!& 
\frac{\rho_i\nu_{in}}{LE_zB_{ext}}
(\partial_y u_y^i)_o^2(1-{\tilde\upsilon})^2
\big[L\Delta_i^3\!+\Delta_iL^3\big]
\nonumber\\
&\!\!\!\approx\!\!\!& 
{\tilde\nu}(1-{\tilde\upsilon})^2\big/(1+{\tilde\rho}{\tilde\upsilon}^2).
\label{Qin_Em_RATIO}
\eeq
Here, to obtain the penultimate expression, we use estimates
$(u_x^i-u_x^n)^2\approx (\partial_x u_x^i-\partial_x u_x^n)_o^2\,x^2
=(\partial_y u_y^i)_o^2(1-{\tilde\upsilon})^2x^2$ and
$(u_y^i-u_y^n)^2\approx (\partial_y u_y^i-\partial_y u_y^n)_o^2\,y^2
=(\partial_y u_y^i)_o^2(1-{\tilde\upsilon})^2y^2$ 
[see eqs.~(\ref{U_N_NORM}), (\ref{DIV_U})]; 
to obtain the final expression, we use inequality $\Delta_i\ll L$, 
eqs.~(\ref{Si_VAi}), (\ref{NU_NORM}), (\ref{Up_Y_Y_EQ}), 
(\ref{DELTA_i}), and we neglect factors of order unity. Note that,
due to eq.~(\ref{UPSILON}), $Q_{in}/{\cal E}_m\lesssim 1$, as one
expects. 
We also see that $Q_{in}/{\cal E}_m$ increases with collisionality for 
small $\tilde\nu$, reaches a peak value that is around unity as 
collisionality increases, and then declines again as 
$\tilde v\rightarrow 1$.

%------------------------------------------------------------------------------------------

\section{Solution}
\label{SEC_SOLUTION}

We solve the nine equations~(\ref{AMPERES_LAW}), 
(\ref{OHMS_LAW_Z_CENTER}), (\ref{Bx_Y}), (\ref{Ez_XX}), 
(\ref{Ez_XY}), (\ref{Up_Y_Y_EQ})-(\ref{V_R}) 
for nine unknowns: $j_o$, $E_z$, $V_R$, $\delta$, 
$\Delta_i$, $(\partial_y u_y^i)_o$, $(\partial_y B_x)_o$, 
${\tilde\upsilon}$ and ${\tilde\gamma}$. 
For the presentation of the solution, it is convenient 
to express resistivity $\eta$ and ion density $\rho_i$ in terms of 
Lundquist number $S_i$, Alfven velocity $V_{Ai}$, field
$B_{ext}$ and scale $L_{ext}$, see eq.~(\ref{Si_VAi}). 
It is also helpful to express the ion charge density $ne$ in terms 
of the ion inertial length $d_i$, 
\beq
\begin{array}{l}
ne=\sqrt{\rho_i}/d_i=B_{ext}/d_iV_{Ai},
\\
d_i=(m_i/ne^2)^{1/2}.
\end{array}
\label{D_i}
\eeq
>From eqs.~(\ref{D}) and~(\ref{UPSILON}), we find that
$D_-=\sqrt{{\tilde\rho}{\tilde\nu}}$ if ${\tilde\nu}\ll 1$, 
and $D_-=-{\tilde\rho}{\tilde\nu}{\tilde\upsilon}$ if 
${\tilde\nu}\gg 1$. Thus, if ${\tilde\nu}\not\approx 1$, 
then condition $D_-\ne 0$ is satisfied, and the solution is 
\beq
&&\!\!\!\!\!\!\!\!\!
{\tilde m} \equiv 1+\min\{{\tilde\nu},{\tilde\rho}\},
\label{RHO_TILDE_S}
\\
&&\!\!\!\!\!\!\!\!\!
{\tilde\nu} \approx \sqrt{{\tilde m}}\,\nu_{in}L/V_{Ai}
\approx{}
\nonumber\\
&&\!\!\!\!\!\!\!\!\!
\hphantom{{\tilde\nu}}\approx 
(\nu_{in}L/V_{Ai})\big[1+\min\{\nu_{in}L/V_{Ai},
\sqrt{{\tilde\rho}}\}\big],
\label{NU_S}
\\
&&\!\!\!\!\!\!\!\!\!
{\tilde\upsilon} \approx 
\min\!\big\{\sqrt{{\tilde\nu}/{\tilde\rho}},1\big\},
\\
&&\!\!\!\!\!\!\!\!\!
{\tilde\gamma} \approx 
\sqrt{{\tilde m}}\, S_id_i^2\big/LL_{ext},
\label{GAMMA_S}
\\
&&\!\!\!\!\!\!\!\!\!
(\partial_yu_y^i)_o \approx 
V_{Ai}\big/L\sqrt{{\tilde m}},
\\
&&\!\!\!\!\!\!\!\!\!
j_o \approx \frac{\sqrt{S_i}B_{ext}}{\sqrt{LL_{ext}}}
\left[\frac{1}{\sqrt{{\tilde m}}}
      +\frac{S_id_i^2}{LL_{ext}}\right]^{1/2}\!\!\!\!,
\\
&&\!\!\!\!\!\!\!\!\!
E_z \approx 
\frac{V_{Ai}B_{ext}\sqrt{L_{ext}}}{\sqrt{S_i}\sqrt{L}}
\left[\frac{1}{\sqrt{{\tilde m}}}
      +\frac{S_id_i^2}{LL_{ext}}\right]^{1/2},
\\
&&\!\!\!\!\!\!\!\!\!
(\partial_yB_x)_o \approx \frac{B_{ext}\sqrt{L_{ext}}}{\sqrt{S_i}\,L^{3/2}}
\left[\frac{1}{\sqrt{{\tilde m}}}
      +\frac{S_id_i^2}{LL_{ext}}\right]^{-1/2}\!\!\!,
\qquad
\\
&&\!\!\!\!\!\!\!\!\!
\delta \approx \frac{\sqrt{LL_{ext}}}{\sqrt{S_i}}
\left[\frac{1}{\sqrt{{\tilde m}}}
      +\frac{S_id_i^2}{LL_{ext}}\right]^{-1/2}\!\!\!\!,
\qquad
\\
&&\!\!\!\!\!\!\!\!\!
\Delta_i \approx \frac{\sqrt{LL_{ext}}}{\sqrt{S_i}}\,
\sqrt{{\tilde m}}
\left[\frac{1}{\sqrt{{\tilde m}}}
      +\frac{S_id_i^2}{LL_{ext}}\right]^{1/2}\!\!,
\label{DELTA_i_S}
\\
&&\!\!\!\!\!\!\!\!\!
V_R \approx E_z/B_{ext}\approx (\Delta_i/L)u_{out}^i, 
\label{V_R_S}
\\
&&\!\!\!\!\!\!\!\!\!
u_{out}^i \approx (\partial_yu_y^i)_oL \approx
V_{Ai}\big/\sqrt{{\tilde m}}.
\label{Ui_OUT_S}
\eeq
Here we define ${\tilde m}$, which is the factor by which 
the ion particle mass is effectively increased due to the ion-neutral 
collisions (as we shall see below). We also introduce $u_{out}^i$, which 
is the ion outflow velocity in the downstream region 
outside the reconnection layer (i.e.~at $x=0$ and $y\approx L$). 
Note that eq.~(\ref{V_R_S}) essentially represents the mass 
conservation law for the ions. 
Let analyze the above solution for the case 
\beq
{\tilde\rho}=\rho_n/\rho_i\gg 1, 
\label{LARGE_TILDE_RHO}
\eeq
which holds for molecular clouds, protostellar disks, and the solar 
chromosphere.\footnote{
The case ${\tilde\rho}\lesssim 1$ is not very interesting because 
in this case ${\tilde m}\approx 1$, the effective ion 
mass is comparable to $m_i$, and the neutrals do not 
significantly influence the reconnection process, see 
eqs.~(\ref{RHO_TILDE_S})-(\ref{Ui_OUT_S}). 
}  

Depending on the value of parameter ${\tilde\nu}$, we have the 
following cases for magnetic reconnection.

The first case is when ${\tilde\nu}\ll 1$, and, as a result, 
${\tilde m}=1$ in eqs.~(\ref{RHO_TILDE_S})-(\ref{Ui_OUT_S}).
In this case the ion-neutral collisions are negligible because their 
frequency is very small compared to the ion inflow and outflow rates, 
$\nu_{in}\ll (\partial_yu_y^i)_o \approx 
V_{Ai}/L\approx u_y^i(L)/L\approx |u_x^i(\Delta_i)|/\Delta_i$.
The ion-neutral coupling is weak.
The neutral particles carry a negligible fraction of the total plasma 
kinetic energy, ${\tilde\rho}{\tilde\upsilon}^2\approx{\tilde\nu}\ll 1$.
Energy dissipation due to the ion-neutral collisions is very small, 
$Q_{in}/{\cal E}_m\ll 1$ in eq.~(\ref{Qin_Em_RATIO}).

The second case is when 
${\tilde\nu}\gg 1$, and, therefore, ${\tilde m}\gg 1$ 
[assuming eq.~(\ref{LARGE_TILDE_RHO}) holds]. In this case the 
ion-neutral collisions significantly influence the reconnection 
process because their frequency is large, 
$\nu_{in}\gg V_{Ai}/L\gg u_y^i(L)/L\approx |u_x^i(\Delta_i)|/\Delta_i$.
The neutral particles carry most of the plasma kinetic energy 
in this case, ${\tilde\rho}{\tilde\upsilon}^2\gg 1$. 
There is significant energy dissipation $Q_{in}\approx{\cal E}_m$
due to ion-neutral collisions if $1\ll{\tilde\nu}\lesssim{\tilde\rho}$
and the ion-neutral coupling is intermediate (${\tilde\upsilon}<1$).
However, if ${\tilde\nu}\gg{\tilde\rho}$ and the ion-neutral coupling is 
strong (${\tilde\upsilon}=1$), this dissipation is negligible, 
$Q_{in}\approx ({\tilde\rho}/{\tilde\nu}){\cal E}_m\ll {\cal E}_m$, 
[see eqs.~(\ref{UPSILON}) and~(\ref{Qin_Em_RATIO})]. 

In the case ${\tilde\nu}\gg 1$, the ion-neutral 
collisions result in an effective increase in the mass $m_i$ of the 
ion particles by factor ${\tilde m}$. 
This is because equations~(\ref{NU_S}), (\ref{GAMMA_S})-(\ref{DELTA_i_S}), 
(\ref{Ui_OUT_S}) can be obtained from the corresponding equations 
in which ${\tilde m}$ is replaced by unity, by making the 
following substitutions: $m_i\to {\tilde m} m_i$, 
$\rho_i\to {\tilde m}\rho_i$, 
$V_{Ai}=B_{ext}/\sqrt{\rho_i}\to V_{Ai}/\sqrt{{\tilde m}}$, 
$S_i=V_{Ai}L_{ext}/\eta\to S_i/\sqrt{{\tilde m}}$ and
$d_i=\sqrt{m_i/ne^2}\to \sqrt{{\tilde m}}\,d_i$.

In the limiting case of strong coupling when ion-neutral collisions are 
extremely frequent, ${\tilde\nu} \gg {\tilde\rho}$ and 
${\tilde m}={\tilde\rho}$, the neutral particles 
are well coupled to the ions and move together, ${\tilde\upsilon}=1$ 
and ${\bf u}^n={\bf u}^i$ (also $Q_{in}\ll {\cal E}_m$). 
In this case the neutrals and ions behave as a single fluid of density 
$\rho_i+\rho_n=(1+{\tilde\rho})\rho_i\approx{\tilde\rho}\rho_i$, 
and $(\partial_yu_y^i)_o\approx V_A/L=V_{Ai}/L\sqrt{{\tilde m}}$.
These theoretical results are in good agreement with recent 
numerical simulations of reconnection in solar chromosphere 
\citep{smith_sakai_2008}, and with previous theoretical studies
\citep{zweibel_1989,zaqarashvili_2011}.

The last case for magnetic reconnection left to consider is when 
${\tilde\nu}\approx 1$. In this case the determinant $D_-$ 
of the system of equations~(\ref{Uz_I_XX_EQ})-(\ref{Uz_N_XX_EQ}) is 
close to zero, and we find from this system that 
$(\partial_{xy}B_z)_o\propto{\tilde\gamma}$ is also close to zero. 
As a result, higher order Taylor expansion terms have to be included 
into our derivations in order to estimate the physical quantities 
inside the reconnection layer in a mathematically rigorous way. 
Fortunately, we do not need to go through these tedious calculations. 
Instead, we note that solution~(\ref{RHO_TILDE_S})-(\ref{Ui_OUT_S}) 
is continuous at ${\tilde\nu}\approx 1$. Therefore, the
case ${\tilde\nu}\approx 1$ is not special, 
eqs.~(\ref{RHO_TILDE_S})-(\ref{Ui_OUT_S}) still hold, and the 
approximate solution in this case is similar to that in the 
weak coupling case ${\tilde\nu}\ll 1$ because in both cases 
${\tilde m}\approx 1$.

Next, depending on the value of the Lundquist number $S_i$, 
there are two distinct reconnection regimes that the
solution~(\ref{RHO_TILDE_S})-(\ref{Ui_OUT_S}) describes. 

First, when $S_i\ll L_{ext}^2/d_i^2\sqrt{{\tilde m}}$ (i.e. when 
$\delta$ computed from the classical Sweet-Parker theory is 
larger than $d_i{\tilde m}^{1/4}$), 
a modified Sweet-Parker reconnection regime takes place, for which
$\Delta_i\approx\delta\approx {\tilde m}^{1/4}L_{ext}/\sqrt{S_i}$,
$L\approx L_{ext}$ (because $\Delta_i\approx\delta$), 
${\tilde\gamma}\approx \sqrt{{\tilde m}}\,S_id_i^2/L_{ext}^2\ll 1$,
$j_o\approx\sqrt{S_i}\,B_{ext}/L_{ext}{\tilde m}^{1/4}$,
$E_z\approx V_{Ai}B_{ext}/\sqrt{S_i}{\tilde m}^{1/4}$, 
$(\partial_yB_x)_o\approx {\tilde m}^{1/4}B_{ext}/L_{ext}\sqrt{S_i}$, 
and the quadrupole field $B_z\approx (\partial_{xy}B_z)_oL\delta 
= ne(\partial_yu_y^i)_o{\tilde\gamma}L\delta \ll B_{ext}$.
The difference between this regime and 
the classical Sweet-Parker reconnection \citep{sweet_1958,parker_1963}
is that in the former the ion particle mass $m_i$ is effectively increased 
by the factor ${\tilde m}$ due to ion-neutral collisions.

Second, there is a Hall reconnection regime when
$S_i\approx L_{ext}^2/d_i^2\sqrt{{\tilde m}}$. 
With the Lundquist number value 
$S_i\approx L_{ext}^2/d_i^2\sqrt{{\tilde m}}$ 
substituted in, equations~(\ref{NU_S})-(\ref{Ui_OUT_S}) give
$\Delta_i\approx d_i\sqrt{{\tilde m}}$,
$\delta\approx \Delta_i(L/L_{ext})\lesssim \Delta_i$,  
$L\lesssim L_{ext}$, 
${\tilde\gamma}\approx L_{ext}/L\gtrsim 1$, 
$j_o\approx B_{ext}L_{ext}/d_iL\sqrt{{\tilde m}}$, 
$E_z\approx V_{Ai}B_{ext}(d_i/L)$ 
\citep[also see][for the  fully-ionized plasma case]{cowley_1985}, 
$(\partial_yB_x)_o\approx \sqrt{{\tilde m}}\,B_{ext}d_i/LL_{ext}$,
and the quadrupole field $B_z\approx (\partial_{xy}B_z)_oL\delta
\approx B_{ext}$ is comparable to the reconnecting field $B_{ext}$.

Unfortunately, our approach does not allow us to calculate the 
reconnection layer length $L$ in the Hall regime. 
However, similar to \citet{malyshkin_2009,malyshkin_2010},
a plausible conjecture can be made that the above Hall 
reconnection regime represents a transition to fast 
collisionless reconnection, during which the reconnection 
(electron) layer thickness $\delta$ decreases from $d_p$ 
to the electron inertial length $d_e$. This conjecture is 
based on numerical simulations, theory, laboratory and space 
observations of magnetic reconnection in fully ionized plasmas
\citep[e.g.,][]{biskamp_1997,shay_1998,pritchett_2001,cassak_2005,
wygant_2005,daughton_2006,drake_2006,daughton_2007,drake_etal_2008,
ji_daughton_2008,yamada_2010}. Note that at the onset of Hall 
reconnection the layer length is $L\approx L_{ext}$; 
the transition to fast collisionless reconnection is accompanied 
by shrinking of $L$ relative to $L_{ext}$.

When the layer thickness $\delta$ reaches $d_e$, electron inertia 
effects become important. In this study we omitted electron inertia, 
and, therefore, we cannot describe this fast reconnection regime, which 
we plan to consider in the future. At the present time, the 
important result for an application to astrophysical systems is that 
the Hall term becomes important in the generalized Ohm's law, the onset
of Hall reconnection occurs, and a transition to fast collisionless 
reconnection happens when 
\beq
&&\!\!\!\!\!
\frac{\delta_{SP}^2}{d_i^2} = \frac{L_{ext}^2}{S_i d_i^2} 
= \frac{L_{ext}\eta}{V_{Ai}d_i^2} \approx
\sqrt{{\tilde m}} = {}
\nonumber\\
&&\!\!\!\!\!
\qquad\qquad\qquad
{}=\sqrt{1+\min\{{\tilde\nu},{\tilde\rho}\}}.
\label{FAST_RECONNECTION_ONSET}
\eeq
Here $\delta_{SP}\equiv L_{ext}/\sqrt{S_i}$ is the classical
Sweet-Parker reconnection layer thickness, and ${\tilde\nu}$ is 
given by eq.~(\ref{NU_S}) with $L\approx L_{ext}$ for the 
onset of Hall reconnection.
It is important that, if ${\tilde\nu}\gg 1$ and ${\tilde\rho}\gg 1$, 
then the Lundquist number value at which the transition to fast 
reconnection occurs is much lower than the corresponding value 
for the fully ionized plasma case ${\tilde\nu}=0$.

Using eq. (\ref{Si}), eq. (\ref{FAST_RECONNECTION_ONSET}) 
can be recast as
\beq
L_{ext} \approx L_{ext}^{Hall}\equiv 
d_i\omega_{ce}\tau_e\sqrt{{\tilde m}}
\label{FAST_RECONNECTION_ONSET_2}
\eeq
for the onset of Hall reconnection. 
Let us now analyze the onset of Hall 
reconnection in terms of the global scale $L_{ext}$. 
Refer to eqs.~(\ref{ETA_TAU_E}),  %(\ref{V_Ai_VALUE_2}), 
(\ref{NU_S}) and~(\ref{FAST_RECONNECTION_ONSET_2}). 
First, note that as the value of $L_{ext}$ decreases from large 
to small, the ion-neutral coupling changes from strong to 
intermediate at 
$L_{ext}\approx L_{ext}^{s\leftrightarrow i}
\equiv\sqrt{{\tilde\rho}}V_{Ai}/\nu_{in}$ 
(when ${\tilde\nu}\approx{\tilde\rho}$), 
and changes to weak coupling at 
$L_{ext}\approx L_{ext}^{i\leftrightarrow w}\equiv V_{Ai}/\nu_{in}$ 
(when ${\tilde\nu}\approx 1$).
Now, we consider how the onset of Hall reconnection depends on whether 
it happens in the strong, intermediate, or weak coupling case. 
If $\nu_{in}\gtrsim\eta/d_i^2=(m_e/m_i)\tau_e^{-1}$, then the
Hall reconnection onset occurs at 
$L_{ext}^{Hall}\gtrsim L_{ext}^{s\leftrightarrow i}$,
when the coupling is strong and ${\tilde m}={\tilde\rho}$.
If $\nu_{in}\lesssim\eta/d_i^2=(m_e/m_i)\tau_e^{-1}$, then the
onset occurs at 
$L_{ext}^{Hall}\lesssim L_{ext}^{i\leftrightarrow w}$,
when the coupling is weak and ${\tilde m}=1$.
Finally, if $\nu_{in}\approx\eta/d_i^2=(m_e/m_i)\tau_e^{-1}$,
then the onset of Hall reconnection happens in a range
$L_{ext}^{i\leftrightarrow w}\lesssim L_{ext}^{Hall}
\lesssim L_{ext}^{s\leftrightarrow i}$,
which is equivalent to 
$V_{Ai}d_i^2/\eta\lesssim L_{ext}^{Hall}
\lesssim \sqrt{{\tilde\rho}}V_{Ai}d_i^2/\eta$. 
Our analysis does not yield a more precise criterion in this case.

%------------------------------------------------------------------------------------------

\section{Discussion}
\label{SEC_DISCUSSION}

Let us apply our results to magnetic reconnection in  
molecular clouds, protostellar disks, and the solar chromosphere.

Using eqs.~(\ref{NU_VALUE})-(\ref{D_i_VALUE}), we make the 
following estimates
\beq
&&\!\!\!\!\!\!\!\!\!
\frac{\nu_{in}L_{ext}}{V_{Ai}}\approx 
0.13\:{\tilde\rho}\;\frac{\sqrt{m_i/m_p}}{1+m_n/m_i}\,
\frac{n_{cm^{-3}}^{3/2}L_{ext,AU}}{B_{ext,\mu G}},
\qquad
\label{NU_TILDE_VALUE}
\\
&&\!\!\!\!\!\!\!\!\!
\frac{\delta_{SP}^2}{d_i^2}=
\frac{L_{ext}^2}{S_id_i^2} \approx
2.2\times 10^6\,\frac{\sqrt{m_p}}{\sqrt{m_i}}\,
\frac{n_{cm^{-3}}^{3/2}L_{ext,AU}}{T_K^{3/2}B_{ext,\mu G}}\,
\quad
\nonumber\\
&&\!\!\!\!\!\!\!\!\!
\qquad \times 
\big[1+ 1.4\!\times\! 10^{-11}{\tilde\rho}(m_i/m_n)\,T_K^2\big].
\label{DELTA_SQ_RATIO_VALUE}
\eeq
Here, as in Section~\ref{SEC_ASTRO_PARAMETERS}, 
characteristic scale $L_{ext}$ is in the astronomical units (AU), 
magnetic field $B_{ext}$ is in microgauss ($\mu G$), electron number 
density $n$ is in $cm^{-3}$, and temperature $T$ is in Kelvins.
Parameter ${\tilde\nu}$ is calculated by substituting $L=L_{ext}$
into eq.~(\ref{NU_S}) and using eq.~(\ref{NU_TILDE_VALUE}) (recall
that $L\approx L_{ext}$ for the onset of Hall reconnection). 

The second column in Table~\ref{TABLE_PARAMETERS} lists the 
typical values of physical parameters in molecular clouds, 
taken from \citet{mckee_1993}, and the corresponding values 
of $d_i$, $S_i$, ${\tilde\rho}$, ${\tilde\nu}$, 
${\tilde\nu}/{\tilde\rho}$, $\eta_{en}/\eta_{ei}$ and 
$\delta_{SP}^2\big/d_i^2\sqrt{{\tilde m}}$. 
>From the values given for ${\tilde\nu}/{\tilde\rho}$ and
$\delta_{SP}^2\big/d_i^2\sqrt{{\tilde m}}$ we see that 
in molecular clouds the ions and neutrals are typically 
strongly coupled, and condition~(\ref{FAST_RECONNECTION_ONSET}) 
for the onset of fast magnetic reconnection can be satisfied.

For the protostellar disks, we assume the following dependence
of physical parameters on the disk radius $r$, which is measured 
in the astronomical units \citep{wardle_2007}. 
The neutral number density is 
$n_n\approx 5.8\times10^{14}\,r_{AU}^{-11/4}\,cm^{-3}$,
the electron number density is $n\approx 10^{-12}\,n_n$, 
the temperature is $T=280\,r_{AU}^{-1/2}\,K$,
the characteristic length $L_{ext}\approx h\approx 
0.03\,r_{AU}^{5/4}\,AU$ (equal to the disk vertical scale $h$), 
and the magnetic field $B\approx 0.2\,r_{AU}^{-5/4}\,G$ 
(based on a theoretical estimation of the angular momentum 
transport in the disk).
Taking $0.1\,AU\lesssim r\lesssim 100\,AU$, we obtain the 
values reported in the third column of Table~\ref{TABLE_PARAMETERS}.
We see that there is a strong ion-neutral coupling at small radii
and intermediate ion-neutral coupling at large radii. 
Condition~(\ref{FAST_RECONNECTION_ONSET}) for onset 
of fast reconnection can again be satisfied.

In the solar chromosphere, on the other hand, the width of the 
Sweet-Parker layer generally far exceeds the ion skin depth, 
meaning that conditions for fast Hall-mediated reconnection are 
unfavorable (see the last column in Table~\ref{TABLE_PARAMETERS}). 
This is due to the relatively high density and high level of 
ionization compared to the other two systems discussed here.

We conclude that fast collisionless magnetic reconnection may 
indeed be possible in partially ionized plasmas in molecular 
clouds and in protostellar disks. 

We thank Fausto Cattaneo, Hantao Ji, Arieh Konigl, Eric Lawrence, 
Masaaki Yamada for useful and stimulating discussions.
This work was supported by the NSF Center for Magnetic Self-Organization in
Laboratory and Astrophysical Plasmas at the Universities of Chicago and
Wisconsin-Madison.

%------------------------------------------------------------------------------------------

\newpage

\phantom{o}

\newpage

\phantom{o}

\begin{table}[t]
\caption{Typical values of physical parameters 
\label{TABLE_PARAMETERS}
}
\smallskip
\begin{tabular}{|l|c|c|c|}
\hline
 &
$\mbox{molecular clouds}$ &
$\mbox{protostellar disks}$ &
$\mbox{solar chromosphere}$
\\
\hline
ions &
${\rm HCO}^+$, $29m_p$ &
${\rm Mg}^+$, $24m_p$ &
${\rm H}^+$, $m_p$
\\
\hline
neutrals &
${\rm H}_2$, $2m_p$ &
${\rm H}_2$, $2m_p$ &
${\rm H}_2$, $2m_p$ 
\\
\hline
$n_{n,cm^{-3}}$ &
$10^3 \,\mbox{{}--{}}\, 10^5$  &
$2\!\times\!10^9 \,\mbox{{}--{}}\, 3\!\times\!10^{17}$ &
$10^{11} \,\mbox{{}--{}}\, 10^{17}$ 
\\
\hline
$n_{cm^{-3}}$ &
$10^{-5}\sqrt{n_n}$ &
$10^{-12}n_n$ &
$10^{11}(1+10^{-30}n_n^2)$ 
\\
\hline
$T\, (K)$ &
$3 \,\mbox{{}--{}}\, 30$ &
$30 \,\mbox{{}--{}}\, 10^3$ &
$6000$ 
\\
\hline
$B_{ext}$ &
$10 \,\mbox{{}--{}}\, 100\,\mu G$ &
$6\!\times\!10^{-4} \,\mbox{{}--{}}\, 4\,G$ &
$1 \,\mbox{{}--{}}\, 10^3\,G$
\\
\hline
$L_{ext}\, (AU)$ &
$10^5 \,\mbox{{}--{}}\, 10^6$ &
$0.001 \,\mbox{{}--{}}\, 10$ &
$10^{-4} \,\mbox{{}--{}}\, 10^{-2}$ 
\\
\hline
$d_i\, (AU)$ &
$10^{-4} \,\mbox{{}--{}}\, 5\!\times\!10^{-4}$ &
$10^{-8} \,\mbox{{}--{}}\, 2\!\times\!10^{-4}$ &
$5\!\times\!10^{-14} \,\mbox{{}--{}}\, 5\!\times\!10^{-12}$
\\
\hline
$S_i$ &
$ 3\!\times\!10^{13} \,\mbox{{}--{}}\, 3\!\times\!10^{15} $ &
$10^3 \,\mbox{{}--{}}\, 10^8$ &
$3\!\times\!10^7 \,\mbox{{}--{}}\, 3\!\times\!10^{10}$
\\
\hline
${\tilde\rho}$ &
$2\!\times\!10^5 \,\mbox{{}--{}}\, 2\!\times\!10^6$ &
$8\!\times\!10^{10}$ &
$1 \,\mbox{{}--{}}\, 10^4$
\\
\hline
${\tilde\nu}$ &
$4\!\times\!10^5 \,\mbox{{}--{}}\, 4\!\times\!10^{10}$ &
$3\!\times\!10^9 \,\mbox{{}--{}}\, 10^{15}$ &
$3\!\times\!10^3 \,\mbox{{}--{}}\, 10^{17}$
\\
\hline
${\tilde\nu}/{\tilde\rho}$ &
$2 \,\mbox{{}--{}}\, 2\!\times\!10^4$ &
$0.04 \,\mbox{{}--{}}\, 10^4$ &
$10^3 \,\mbox{{}--{}}\, 10^{15}$
\\
\hline
$\eta_{en}/\eta_{ei}$ &
$ 0.004 \,\mbox{{}--{}}\, 0.04 $ &
$10^4 \,\mbox{{}--{}}\, 10^7$ &
$5\!\times\!10^{-4} \,\mbox{{}--{}}\, 3$
\\
\hline
$\delta_{SP}^2\big/d_i^2\sqrt{{\tilde m}}$ &
$0.03 \,\mbox{{}--{}}\, 10^3$ &
$0.001 \,\mbox{{}--{}}\, 50$ &
$500 \,\mbox{{}--{}}\, 10^{14}$
\\
\hline
\end{tabular}
\end{table}

\phantom{o}

\newpage

%------------------------------------------------------------------------------------------

\appendix

\section{Equations with the electron-neutral collisions included}
\label{APPENDIX_A}

In this appendix we prove that, if $m_e\nu_{en}\ll m_i\nu_{in}$,
then the terms that are related to the electron-neutral collisions 
and are proportional to $\nu_{en}$ can be neglected in the reconnection 
equations, except in eq.~(\ref{RESISTIVITY}). While some of these 
terms are clearly small, for example $\rho_e\nu_{en}\ll\rho_i\nu_{in}$ in 
eqs.~(\ref{MOMENTUM_I_ORIGINAL}) and~(\ref{MOMENTUM_N_ORIGINAL}), it is 
not immediately clear that other terms can be omitted. A rigorous 
proof requires deriving general equations with electron-neutral 
collisions included, and then showing that these terms are negligible. 
The general equations can be useful for the case when 
electron-neutral collisions are significant, as long as electron 
inertia can be neglected.\footnote{
Note that electron-neutral collisions result in an effective increase
in the electron mass $m_e$, similar to the increase in $m_i$ due to 
ion-neutral collisions (see Section~\ref{SEC_SOLUTION}).
}
To save space, we do not give all details of the derivations, which 
are tedious but straightforward to carry out along exactly the same 
guidelines that are thoroughly described in the main text. 
Instead we give only the key formulas and results. 
Also, below we prove that the last term in eq.~(\ref{Ez_XY_INITIAL}) 
can be neglected if $D_-\ne 0$.

For brevity of notation, we assume that spatial derivatives are
to be calculated with respect to all indices that are listed after the
comma signs in the subscripts, e.g.~$u_{y,y}^p\equiv\partial_y u_y^p$ and 
$B_{z,xy}\equiv\partial_{xy} B_z$. 

With the electron-neutral collisions included into derivation, it is 
convenient to replace eq.~(\ref{NU_NORM}) by 
\beq
\!\!\!\!\!\!&&
\:{\tilde\nu} \equiv {\tilde\nu}_i+{\tilde\nu}_e,
\qquad
{\tilde\nu}_i \equiv \nu_{in}\left/2(u_{y,y}^i)_o \right.,
\qquad
{\tilde\nu}_e \equiv (m_e/m_i)\nu_{en}\left/2(u_{y,y}^i)_o \right.,
\label{A_NU}
\\
\!\!\!\!\!\!&&
{\tilde\nu}_e/{\tilde\nu}\approx {\tilde\nu}_e/{\tilde\nu}_i\ll 1,
\qquad\qquad
{\tilde\nu}_e{\tilde\gamma}/{\tilde\nu}
\approx {\tilde\nu}_e{\tilde\gamma}/{\tilde\nu}_i\ll 1.
\label{A_SMALL_TERMS_0}
\eeq
Here, ${\tilde\nu}_e/{\tilde\nu}_i\ll 1$ follows directly from 
$m_e\nu_{en}\ll m_i\nu_{in}$, while 
${\tilde\nu}_e{\tilde\gamma}/{\tilde\nu}_i\ll 1$
can be used because we are interested in the conditions 
for a transition to fast collisionless reconnection at 
${\tilde\gamma}\approx 1$ (see Section~\ref{SEC_SOLUTION}).

With the electron-neutral collisions included, eqs.~(\ref{Si_VAi}), 
(\ref{DENSITY_NORM}), (\ref{U_N_NORM})-(\ref{AMPERES_LAW}), (\ref{DELTA_i}) 
and~(\ref{V_R}) are unchanged. Equation~(\ref{OHMS_LAW_Z}) becomes
\beq
E_z &\!\!=\!\!& \eta j_z-(u_x^i-j_x/ne)B_y+(u_y^i-j_y/ne)B_x
-(m_e\nu_{en}/e)(u_z^i-u_z^n)=\mbox{constant}.
\nonumber
\eeq
We see that equation~(\ref{OHMS_LAW_Z_CENTER}) stays the same. 
Equations~(\ref{Uz_I_XX})-(\ref{Uz_N_YY}) become
\beq
(u_{z,xx}^i)_o &\!\!=\!\!&
-{\tilde\gamma}(1+{\tilde\nu}{\tilde\rho}{\tilde\upsilon}/D_-)
(e/m_i)(B_{y,x})_o 
-({\tilde\nu}_e{\tilde\rho}{\tilde\upsilon}/D_-)(1/ne)(j_{z,xx})_o,
\nonumber
\\
(u_{z,xx}^n)_o &\!\!=\!\!&
{\tilde\gamma}({\tilde\nu}/D_-)(e/m_i)(B_{y,x})_o 
+({\tilde\nu}_e/D_-)(1/ne)(j_{z,xx})_o,
\nonumber
\\
(u_{z,yy}^i)_o &\!\!=\!\!&
{\tilde\gamma}(1-{\tilde\nu}{\tilde\rho}{\tilde\upsilon}/D_+)
(e/m_i)(B_{x,y})_o 
+({\tilde\nu}_e{\tilde\rho}{\tilde\upsilon}/D_+)(1/ne)(j_{z,yy})_o,
\nonumber
\\
(u_{z,yy}^n)_o &\!\!=\!\!&
{\tilde\gamma}({\tilde\nu}/D_+)(e/m_i)(B_{x,y})_o 
-({\tilde\nu}_e/D_+)(1/ne)(j_{z,yy})_o,
\nonumber
\eeq
where $D_\pm$ is still given by eq.~(\ref{D}) with ${\tilde\nu}$ defined
by eq.~(\ref{A_NU}) now. We assume that $D_-$ is non-zero 
(the case $D_-=0$ is discussed in Section~\ref{SEC_SOLUTION}). 
Equations~(\ref{Ez_XX_INITIAL}) and~(\ref{Ez_YY_INITIAL}) become
\beq
{} -[\eta+\eta_{en}{\tilde\nu}_e(1+{\tilde\rho}{\tilde\upsilon})/D_-]
(j_{z,xx})_o = 2(u_{y,y}^i)_o(B_{y,x})_o
[1+{\tilde\gamma}(1+{\tilde\nu}_e{\tilde\rho}{\tilde\upsilon}/D_-)],
\label{A_Ez_XX_INITIAL}
\\
{} -[\eta-\eta_{en}{\tilde\nu}_e(1+{\tilde\rho}{\tilde\upsilon})/D_+]
(j_{z,yy})_o = 2(u_{y,y}^i)_o(B_{x,y})_o
[1+{\tilde\gamma}(1-{\tilde\nu}_e{\tilde\rho}{\tilde\upsilon}/D_+)].
\eeq
Taking the ratio of these two equations, we obtain a general version
of eq.~(\ref{Bx_Y}): 
\beq
(B_{x,y})_o &\!\!\!\!\approx\!\!\!\!&
\frac{B_{ext}\delta}{L^2}\;
\frac{1+{\tilde\gamma}(1+{\tilde\nu}_e{\tilde\rho}{\tilde\upsilon}/D_-)}
{1+{\tilde\gamma}(1-{\tilde\nu}_e{\tilde\rho}{\tilde\upsilon}/D_+)}\;
\frac{\eta-\eta_{en}{\tilde\nu}_e(1+{\tilde\rho}{\tilde\upsilon})/D_+}
{\eta+\eta_{en}{\tilde\nu}_e(1+{\tilde\rho}{\tilde\upsilon})/D_-}.
\label{A_Bx_Y}
\eeq
We can rewrite eq.~(\ref{A_Ez_XX_INITIAL}) in an approximate form that 
corresponds to eq.~(\ref{Ez_XX}),
\beq
{} [\eta+\eta_{en}{\tilde\nu}_e(1+{\tilde\rho}{\tilde\upsilon})/D_-]
j_o^2 \approx B_{ext}^2(u_{y,y}^i)_o
[1+{\tilde\gamma}(1+{\tilde\nu}_e{\tilde\rho}{\tilde\upsilon}/D_-)]. 
\label{A_Ez_XX}
\eeq
Next, with the electron-neutral collisions included, 
eq.~(\ref{Ez_XY_INITIAL}) becomes
\begin{eqnarray}
0 &\!\!=\!\!&
-\eta\left[(\partial_{xx}+\partial_{yy})(B_{z,xy})\right]_o
-(m_e\nu_{en}/e)
[(\partial_{xx}+\partial_{yy})(u_{y,y}^i-u_{y,y}^n)]_o
\nonumber\\
&\!\!\!\!& +[(j_{z,xx})_o(B_{x,y})_o
(1+{\tilde\nu}_e{\tilde\rho}{\tilde\upsilon}/D_-)
+(j_{z,yy})_o(B_{y,x})_o
(1-{\tilde\nu}_e{\tilde\rho}{\tilde\upsilon}/D_+)]/ne
\nonumber\\
&\!\!\!\!& 
+(2{\tilde\gamma}{\tilde\nu}{\tilde\rho}^2{\tilde\upsilon}^2/D_-D_+)
(e/m_i)(B_{y,x})_o(B_{x,y})_o \approx {}
\nonumber
\\
{}&\!\!\approx\!\!& 
[\eta{\tilde\gamma}+\eta_{en}(1-{\tilde\upsilon})]
ne(u_{y,y}^i)_o/\delta^2
-[(j_o/\delta^2)(B_{x,y})_o
(1+{\tilde\nu}_e{\tilde\rho}{\tilde\upsilon}/D_-)
+(j_o^2/L^2)(1-{\tilde\nu}_e{\tilde\rho}{\tilde\upsilon}/D_+)]/ne
\nonumber\\
&\!\!\!\!& + ({\tilde\gamma}{\tilde\nu}{\tilde\rho}^2{\tilde\upsilon}^2/D_-D_+)
(e/m_i)j_o(B_{x,y})_o.
\label{A_Ez_XY}
\end{eqnarray}
Finally, eqs.~(\ref{UPSILON_EQ}) and~(\ref{Up_Y_Y_EQ}) become
\beq
&&\!\!\!\!\!\!\!
{\tilde\rho}{\tilde\upsilon}^2 =
2{\tilde\nu}(1-{\tilde\upsilon})+2{\tilde\nu}_e{\tilde\gamma},
\label{A_UPSILON_EQ}
\\
&&\!\!\!\!\!\!\!
\rho_i(u_{y,y}^i)_o^2(1+{\tilde\rho}{\tilde\upsilon}^2) \approx
B_{ext}^2/L^2 +j_o(B_{x,y})_o
\label{A_Up_Y_Y_EQ}
\eeq
respectively [also refer to eqs.~(\ref{B_Pe_Pi_YY}) and~(\ref{B_Pn_YY})].
The parameter ${\tilde\upsilon}$ must be non-negative, therefore, 
the physically correct solution of eq.~(\ref{A_UPSILON_EQ}) is
\beq
{\tilde\upsilon}=(1/{\tilde\rho})
\big[\sqrt{{\tilde\nu}^2+2{\tilde\rho}({\tilde\nu}+{\tilde\nu}_e{\tilde\gamma})}
-{\tilde\nu}\big].
\label{A_UPSILON}
\eeq

Next, let us use inequality ${\tilde\nu}_e/{\tilde\nu}_i\ll 1$ or, 
equivalently, ${\tilde\nu}_e/{\tilde\nu}\ll 1$
to simplify the above equations. We still assume that 
$D_-$, given by eq.~(\ref{D}), is not close to zero, and, therefore, 
$|D_\pm|\approx {\tilde\rho}{\tilde\upsilon}
+{\tilde\nu}(1+{\tilde\rho}{\tilde\upsilon})$. We have 
\beq
\frac{{\tilde\nu}_e{\tilde\rho}{\tilde\upsilon}}{|D_\pm|}
\approx \frac{{\tilde\nu}_e{\tilde\rho}{\tilde\upsilon}}
{{\tilde\rho}{\tilde\upsilon}+{\tilde\nu}(1+{\tilde\rho}{\tilde\upsilon})}
\le \frac{{\tilde\nu}_e}{{\tilde\nu}}\ll 1,
\qquad
\frac{\eta_{en}{\tilde\nu}_e(1+{\tilde\rho}{\tilde\upsilon})}{\eta\,|D_\pm|}
\lesssim \frac{{\tilde\nu}_e(1+{\tilde\rho}{\tilde\upsilon})}
{{\tilde\rho}{\tilde\upsilon}+{\tilde\nu}(1+{\tilde\rho}{\tilde\upsilon})}
\lesssim \frac{{\tilde\nu}_e}{{\tilde\nu}}\ll 1.
\label{A_SMALL_TERMS_1}
\eeq
As a result of inequalities~(\ref{A_SMALL_TERMS_0}) 
and~(\ref{A_SMALL_TERMS_1}), 
equations~(\ref{A_Bx_Y}), (\ref{A_Ez_XX}), (\ref{A_Up_Y_Y_EQ}) 
and~(\ref{A_UPSILON}) reduce to 
equations~(\ref{Bx_Y}), (\ref{Ez_XX}), (\ref{Up_Y_Y_EQ}) 
and~(\ref{UPSILON}) respectively. Henceforth, we can use the latter 
instead of the former, except for calculations of $1-{\tilde\upsilon}$ 
for which the more accurate equation~(\ref{A_UPSILON}) should be used 
when $1-{\tilde\upsilon}\ll 1$.

Now, the only proof left to do is to show that eq.~(\ref{A_Ez_XY}) 
reduces to eq.~(\ref{Ez_XY}). This proof is done as follows. 
Using eqs.~(\ref{Si_VAi}), (\ref{AMPERES_LAW}), (\ref{Bx_Y}) 
and~(\ref{A_SMALL_TERMS_1}), we rewrite eq.~(\ref{A_Ez_XY}) as
\beq
\!\!\!\!\!\!&&
\eta{\tilde\gamma}(u_{y,y}^i)_o
+\eta_{en}(1-{\tilde\upsilon})(u_{y,y}^i)_o
- B_{ext}^2/n^2e^2L^2 
+ ({\tilde\gamma}{\tilde\nu}{\tilde\rho}^2{\tilde\upsilon}^2/D_-D_+)
(B_{ext}^2V_{Ai}^2/L^2j_o^2) \approx 0.
\quad
\label{A_Ez_XY_SIMPLER}
\end{eqnarray}
Let us estimate the ratio of the last term and the first term on the 
left-hand-side of eq.~(\ref{A_Ez_XY_SIMPLER}),
\beq
&& \!\!\!\!\!\!\!\! 
\frac{{\tilde\nu}{\tilde\rho}^2{\tilde\upsilon}^2}{|D_-|D_+}
\frac{B_{ext}^2V_{Ai}^2}{\eta L^2j_o^2(u_{y,y}^i)_o} 
\approx \frac{{\tilde\nu}{\tilde\rho}^2{\tilde\upsilon}^2}{|D_-|D_+}
\frac{1+{\tilde\rho}{\tilde\upsilon}^2}{1+{\tilde\gamma}} 
\le \frac{{\tilde\nu}}{(1+{\tilde\nu})^2}
\frac{1+{\tilde\rho}{\tilde\upsilon}^2}{1+{\tilde\gamma}} 
\le \frac{{\tilde\nu}}{(1+{\tilde\nu})^2}
+\frac{{\tilde\rho}{\tilde\upsilon}^2}{{\tilde\nu}(1+{\tilde\gamma})}
\lesssim 1.
\qquad
\nonumber
\eeq
Here we use eqs.~(\ref{Ez_XX}) and~(\ref{Up_Y_Y_EQ})
to obtain the second expression; we use an estimate 
$|D_\pm|\approx {\tilde\rho}{\tilde\upsilon}
+{\tilde\nu}(1+{\tilde\rho}{\tilde\upsilon})
\ge{\tilde\rho}{\tilde\upsilon}(1+{\tilde\nu})$ to obtain the third 
expression (assuming $D_-\ne0$); and we use 
estimates ${\tilde\nu}<(1+{\tilde\nu})^2$, $1+{\tilde\gamma}\ge 1$ 
and ${\tilde\rho}{\tilde\upsilon}^2\lesssim{\tilde\nu}$ 
[see eq.~(\ref{UPSILON})] to obtain the final result. 
Thus, the last term on the left-hand-side of eq.~(\ref{A_Ez_XY_SIMPLER})
can be neglected because it is comparable to the first term or smaller.  

Next let us prove that the second term, 
$\eta_{en}(1-{\tilde\upsilon})(u_{y,y}^i)_o$, can be 
neglected in eq.~(\ref{A_Ez_XY_SIMPLER}) as well. First, the ratio of 
this term and the term $B_{ext}^2/n^2e^2L^2$ is 
\beq
\!\!\!\!\!\!&&
\frac{n^2e^2\eta_{en}(1-{\tilde\upsilon})(u_{y,y}^i)_oL^2}{B_{ext}^2} 
= \frac{\rho_im_e\nu_{en}}{m_i} \frac{(1-{\tilde\upsilon})(u_{y,y}^i)_oL^2}{B_{ext}^2}
= \frac{2{\tilde\nu}_e(1-{\tilde\upsilon})(u_{y,y}^i)_o^2L^2}{V_{Ai}^2} 
\ll\frac{{\tilde\nu}(1-{\tilde\upsilon})}{1+{\tilde\rho}{\tilde\upsilon}^2}.
\label{A_SMALL_TERMS_3}
\eeq
Here we use eqs.~(\ref{RHO}) and~(\ref{RESISTIVITY}) to obtain 
the second expression; 
we use eqs.~(\ref{Si_VAi}) and~(\ref{A_NU}) to obtain the third
expression; and we use eq.~(\ref{Up_Y_Y_EQ}) and inequality 
${\tilde\nu}_e\ll{\tilde\nu}$ to obtain the final result. 
Second, the ratio of the second and the first terms in 
eq.~(\ref{A_Ez_XY_SIMPLER}) is
\beq
\frac{\eta_{en}(1-{\tilde\upsilon})(u_{y,y}^i)_o}
{\eta{\tilde\gamma}(u_{y,y}^i)_o}
=\frac{\eta_{en}}{\eta}\frac{1-{\tilde\upsilon}}{{\tilde\gamma}}\le
\frac{1-{\tilde\upsilon}}{{\tilde\gamma}}.
\label{A_SMALL_TERMS_4}
\eeq

In the case 
${\tilde\nu}\lesssim{\tilde\rho}(1+{\tilde\nu}_e{\tilde\gamma})/{\tilde\nu}$,
from eq.~(\ref{A_UPSILON}) we find 
${\tilde\upsilon}
\approx\sqrt{({\tilde\nu}+{\tilde\nu}_e{\tilde\gamma})/{\tilde\rho}}$
and $1-{\tilde\upsilon}\approx 1$. Therefore, 
${\tilde\nu}(1-{\tilde\upsilon})/(1+{\tilde\rho}{\tilde\upsilon}^2)\approx 
{\tilde\nu}/(1+{\tilde\nu}+{\tilde\nu}_e{\tilde\gamma})\lesssim 1$ in 
eq.~(\ref{A_SMALL_TERMS_3}). Thus, in this case the term 
$\eta_{en}(1-{\tilde\upsilon})(u_{y,y}^i)_o$ can be neglected
in eq.~(\ref{A_Ez_XY_SIMPLER}) because this term is small in comparison with 
the term $B_{ext}^2/n^2e^2L^2$. 

In case 
${\tilde\nu}\gg{\tilde\rho}(1+{\tilde\nu}_e{\tilde\gamma})/{\tilde\nu}$, 
from eq.~(\ref{A_UPSILON}) we obtain 
${\tilde\upsilon}\approx 1$ and $1-{\tilde\upsilon}\approx 
-{\tilde\nu}_e{\tilde\gamma}/{\tilde\nu}+{\tilde\rho}/{\tilde\nu}$. 
We substitute 
$1-{\tilde\upsilon}\approx -{\tilde\nu}_e{\tilde\gamma}/{\tilde\nu}$
into eq.~(\ref{A_SMALL_TERMS_4}) and find 
$|1-{\tilde\upsilon}|/{\tilde\gamma}\approx {\tilde\nu}_e/{\tilde\nu}\ll 1$.
We also substitute $1-{\tilde\upsilon}\approx {\tilde\rho}/{\tilde\nu}$
into eq.~(\ref{A_SMALL_TERMS_3}) and obtain 
${\tilde\nu}(1-{\tilde\upsilon})/(1+{\tilde\rho}{\tilde\upsilon}^2)\approx 
{\tilde\rho}/(1+{\tilde\rho})<1$. Therefore, the term
$\eta_{en}(1-{\tilde\upsilon})(u_{y,y}^i)_o$ can again be neglected
in eq.~(\ref{A_Ez_XY_SIMPLER}) because this term is small in comparison 
with either the term $\eta{\tilde\gamma}(u_{y,y}^i)_o$ or the term
$B_{ext}^2/n^2e^2L^2$.

As a result of the above estimates, the second and the fourth terms 
in eq.~(\ref{A_Ez_XY_SIMPLER}) can be omitted. Therefore, this equation
and eq.~(\ref{A_Ez_XY}) reduce to eq.~(\ref{Ez_XY}).

%------------------------------------------------------------------------------------------

\section{Derivation of equations~(\ref{Pe_Pp_YY}) and~(\ref{Pn_YY})}
\label{APPENDIX_B}

As in the previous appendix, here we include electron-neutral 
collisions, and, to save space, we again assume 
that spatial derivatives are to be calculated with respect to all 
indexes listed after the comma signs in the subscripts, 
e.g.~$u_{y,y}^p\equiv\partial_y u_y^p$ and 
$B_{z,xy}\equiv\partial_{xy} B_z$.

Our derivation of equations~(\ref{Pe_Pp_YY}) and~(\ref{Pn_YY}), to some 
degree, is similar to the Sweet-Parker arguments for the pressure 
drop along and across the reconnection layer. To be precise, 
we integrate pressure gradient vectors along the rectangular contour 
$O \to M \to {\tilde M} \to {\tilde O}$ shown in Figure~\ref{FIGURE_LAYER} 
and use the force balance condition for the plasma slowly 
inflowing across the reconnection layer. 
Note that point $O$ is the reconnection layer center, and point 
$M$ is in the upstream region at the edge of the ion layer at
$x\approx\Delta_i$ and $y=0$. Let ${\tilde y}$ be the 
y-coordinate of points ${\tilde{\rm M}}$ and ${\tilde{\rm O}}$. 
We proceed as follows.

We consider the limit when points ${\tilde O}$ and ${\tilde M}$
are infinitesimally close to point $O$ and $M$ respectively, and, therefore, 
${\tilde y}\to+0$. For infinitesimally small values of the y-coordinate, 
we use Taylor expansions in the $y$ coordinate for the $x$- 
and $y$-components of the velocities, current and magnetic field, 
\beq
\begin{array}{l l}
u_x^i=u_x^{i,(0)}(x)+(y^2/2)u_{x,yy}^{i,(0)}(x),
&
u_y^i=yu_{y,y}^{i,(0)}(x)+(y^3/6)u_{y,yyy}^{i,(0)}(x),
\\
u_x^n=u_x^{n,(0)}(x)+(y^2/2)u_{x,yy}^{n,(0)}(x),
&
u_y^n=yu_{y,y}^{n,(0)}(x)+(y^3/6)u_{y,yyy}^{n,(0)}(x),
\\
j_x=j_x^{(0)}(x)+(y^2/2)j_{x,yy}^{(0)}(x),
&
j_y=yj_{y,y}^{(0)}(x)+(y^3/6)j_{y,yyy}^{(0)}(x),
\\
B_x=yB_{x,y}^{(0)}(x)+(y^3/6)B_{x,yyy}^{(0)}(x),
&
B_y = B_y^{(0)}(x)+(y^2/2)B_{y,yy}^{(0)}(x).
\end{array}
\label{APP_A_EXPANSIONS}
\eeq
Here the variables with the superscripts $\,^{(0)}$ are calculated at $y=0$ 
and depend only on coordinate $x$. 

Neglecting time derivatives for a quasi-stationary case, we rewrite 
the momentum eqs.~(\ref{MOMENTUM_I}) and~(\ref{MOMENTUM_N}) as
\begin{eqnarray}
{\bf\nabla}(P_e+P_i+B^2/2) &\!\!=\!\!&
-\rho_i({\bf u}^i{\bf\nabla}){\bf u}^i + ({\bf B}{\bf\nabla}){\bf B}
-(\rho_i\nu_{in}+\rho_e\nu_{en})({\bf u}^i-{\bf u}^n)+(m_e\nu_{en}/e){\bf j},
\qquad
\label{APP_A_Vi_EQUATION}
\\
{\bf\nabla}P_n &\!\!=\!\!&
-\rho_n({\bf u}^n{\bf\nabla}){\bf u}^n 
+(\rho_i\nu_{in}+\rho_e\nu_{en})({\bf u}^i-{\bf u}^n)-(m_e\nu_{en}/e){\bf j}.
\qquad
\label{APP_A_Vn_EQUATION}
\end{eqnarray}
Next, we calculate the line integrals of all terms in these 
two equations along the contour $O \to M \to {\tilde M} \to {\tilde O}$ 
(see Fig.~\ref{FIGURE_LAYER}). 
In these calculations we use Taylor expansions~(\ref{APP_A_EXPANSIONS}), 
and we keep only the terms up to the leading, second order in ${\tilde y}$ 
(because ${\tilde y}\to+0$). As a result, we obtain
\beq
\int[{\bf\nabla}(P_e+P_i+B^2/2)]{\bf dl} &\!\!=\!\!& 
[P_e(0,{\tilde y})+P_i(0,{\tilde y})+B_x^2(0,{\tilde y})/2]-[P_e(0,0)+P_i(0,0)]=
\nonumber\\
&\!\!=\!\!& ({\tilde y}^2/2)\big[(P_{e,yy}+P_{i,yy})_o+(B_{x,y})_o^2\big],
\label{APP_A_INTEGRALS_1}
\\
\int[{\bf\nabla}P_n]{\bf dl} &\!\!=\!\!& 
P_n(0,{\tilde y})-P_n(0,0) = ({\tilde y}^2/2)(P_{n,yy})_o,
\\
\int[({\bf u}^i{\bf\nabla}){\bf u}^i]{\bf dl} &\!\!=\!\!&
({\tilde y}^2/2) \Bigl[\bigl(u_{x,x}^{i,(m)}\bigr)^2
-u_x^{i,(m)}\bigl(u_{x,xx}^{i,(m)}+u_{x,yy}^{i,(m)}\bigr)
+2\int_O^M u_{x,x}^{i,(0)}u_{x,yy}^{i,(0)}\,dx\Bigr]=
\nonumber\\
&\!\!=\!\!& ({\tilde y}^2/2)\,
O\big\{(u_{y,y}^i)_o^2\Delta_i^2/L^2\big\}
=({\tilde y}^2/2)\, o\big\{(u_{y,y}^i)_o^2\big\},
\label{APP_A_INTEGRALS_2}
\\
\int[({\bf u}^n{\bf\nabla}){\bf u}^n]{\bf dl} 
&\!\!=\!\!& ({\tilde y}^2/2)\, o\big\{(u_{y,y}^n)_o^2\big\}, 
\quad\mbox{by analogy with eq.~(\ref{APP_A_INTEGRALS_2})},
\\
\int[({\bf B}{\bf\nabla}){\bf B}]{\bf dl} &\!\!=\!\!&
({\tilde y}^2/2) \Bigl[(B_{x,y})_o^2 
+B_{y}^{(m)}B_{y,yy}^{(m)}+B_{x,y}^{(m)}j_{z}^{(m)}
-\int_O^M \bigl(B_{y}^{(0)}B_{x,y}^{(0)}\bigr)_{\!,yy}\:dx\,\Bigr]=
\nonumber\\
&\!\!=\!\!& ({\tilde y}^2/2) \big[ (B_{x,y})_o^2 
+B_{y}^{(m)}B_{y,yy}^{(m)}+O\big\{j_o(B_{x,y})_o\delta/L\big\}\big]=
\nonumber\\
&\!\!=\!\!& ({\tilde y}^2/2)\big[(B_{x,y})_o^2 
+B_{y}^{(m)}B_{y,yy}^{(m)}+o\big\{j_o(B_{x,y})_o\big\}\big],
\\
\int({\bf u}^i-{\bf u}^n){\bf dl} &\!\!=\!\!& 
-({\tilde y}^2/2) \Bigl[u_{x,x}^{i,(m)}-u_{x,x}^{n,(m)}
+\int_O^M \bigl(u_{x,yy}^{i,(0)}-u_{x,yy}^{n,(0)}\bigr)\,dx\Bigr]
\nonumber\\
&\!\!=\!\!& 
({\tilde y}^2/2)\, O\big\{(u_{x,x}^i-u_{x,x}^n)_o\Delta_i/L\big\}=
({\tilde y}^2/2)\, o\big\{(u_{y,y}^i-u_{y,y}^n)_o\big\},
\\
\int{\bf j}\,{\bf dl} &\!\!=\!\!&
-({\tilde y}^2/2) \Bigl[j_{x,x}^{(m)}+\int_O^M j_{x,yy}^{(0)}\,dx\Bigr]
\nonumber\\
&\!\!=\!\!& 
({\tilde y}^2/2)\, O\big\{(B_{z,xy})_o\delta/L\big\}=
({\tilde y}^2/2)\, o\big\{(j_{y,y})_o\big\}.
\label{APP_A_INTEGRALS_3}
\eeq
Here, the variables inside parentheses $(...)_o$ are evaluated 
at the central point $O$; the variables with the superscript $\,^{(m)}$ 
are calculated at point M; notation $O\{...\}$ denotes terms that, 
in absolute value, are comparable to the terms inside the brackets 
$\{...\}$; and notation $o\{...\}$ denotes terms that are small 
compared to the terms inside the brackets.
In derivations of eqs.~(\ref{APP_A_INTEGRALS_1})-(\ref{APP_A_INTEGRALS_3})
we use equations $u_{x,x}^i=-u_{y,y}^i$, $u_{x,x}^n=-u_{y,y}^n$,  
$j_{x,x}=-j_{y,y}=B_{z,xy}$. To derive the final approximate expressions 
in eqs.~(\ref{APP_A_INTEGRALS_1})-(\ref{APP_A_INTEGRALS_3}), we use
the following estimates at point $M$ (which is at the ion layer edge):
$\partial_y\approx 1/L$, $\partial_x\approx 1/L$, 
$u_x^i\approx(u_{x,x}^i)_o\Delta_i=-(u_{y,y}^i)_o\Delta_i$, 
$u_x^n\approx-(u_{y,y}^n)_o\Delta_i$, 
$B_y\approx B_{ext}\approx j_o\delta$, 
$B_{x,y}\approx(B_{x,y})_o$, $j_z\approx B_{ext}/L$,
$j_x\approx(j_{x,x})_o\delta=(B_{z,xy})_o\delta$ (note that, in the Hall regime, 
the Hall term supports $E_z$ outside of the electron layer), and 
$\delta\lesssim\Delta_i\ll L$ for a slow reconnection in a thing layer.
We also use the following estimates for the integral terms: 
$\int_O^M u_{x,x}^{i,(0)}u_{x,yy}^{i,(0)}\,dx 
\approx \int_O^M (u_{x,x}^i)_o^2\, x L^{-2}dx
\approx (u_{y,y}^i)_o^2\Delta_i^2/L^2$,
$\int_O^M (B_{y}^{(0)}B_{x,y}^{(0)})_{,yy}dx
\approx \int_O^M (B_{x,y})_oB_{ext} L^{-2}dx
\approx (B_{x,y})_oj_o\delta\,\Delta/L^2$,
$\int_O^M (u_{x,yy}^{i,(0)}-u_{x,yy}^{n,(0)})\,dx
\approx \int_O^M (u_{x,x}^i-u_{x,x}^n)_o\, x L^{-2}dx
\approx -(u_{y,y}^i-u_{y,y}^n)_o\Delta_i^2/L^2$, and
$\int_O^M j_{x,yy}^{(0)} dx
\approx \int_O^M (j_{x,x})_o\delta\,L^{-2}dx
\approx -(B_{z,xy})_o\delta\Delta_i/L^2$.

Taking the line integrals of eqs.~(\ref{APP_A_Vi_EQUATION}) 
and~(\ref{APP_A_Vn_EQUATION}), using 
eqs.~(\ref{APP_A_INTEGRALS_1})-(\ref{APP_A_INTEGRALS_3}) and
an estimate $B_{y}^{(m)}B_{y,yy}^{(m)}\approx -B_{ext}^2/L^2$, 
we obtain
\beq
(P_{e,yy}+P_{i,yy})_o &\!\!\approx\!\!&
-B_{ext}^2/L^2 +o\big\{\rho_i(u_{y,y}^i)_o^2\big\}
+o\big\{j_o(B_{x,y})_o\big\}
+o\big\{(m_e\nu_{en}/e)(j_{y,y})_o\big\}
\nonumber\\
&\!\!\!\!& +o\big\{(\rho_i\nu_{in}+\rho_e\nu_{en})
(u_{y,y}^i-u_{y,y}^n)_o \big\},
\label{B_Pe_Pi_YY}
\\
(P_{n,yy})_o &\!\!=\!\!&
o\big\{\rho_n(u_{y,y}^n)_o^2\big\}
+o\big\{(m_e\nu_{en}/e)(j_{y,y})_o\big\}
+o\big\{(\rho_i\nu_{in}+\rho_e\nu_{en})
(u_{y,y}^i-u_{y,y}^n)_o \big\}.  
\qquad
\label{B_Pn_YY}
\eeq
These equations reduce to eqs.~(\ref{Pe_Pp_YY}) and~(\ref{Pn_YY})
when the terms associated with the electron-neutral collisions and 
proportional to $\nu_{en}$ are neglected. Note that the term
$B_{y}^{(m)}B_{y,yy}^{(m)}\approx -B_{ext}^2/L^2$ represents the
drop of the outside magnetic pressure along the reconnection layer.

%------------------------------------------------------------------------------------------

%------------------------------------------------------------------------------------------

%\clearpage

\end{document}